\documentclass[reprint, nofootinbib,noshowpacs,superscriptaddress,11pt]{revtex4}
\usepackage{blindtext}
\usepackage{scrextend}

\usepackage{footnote}
\usepackage{slashed}
\usepackage{lipsum}
\usepackage[colorlinks=true,allcolors=blue]{hyperref}
\usepackage{amssymb,amsmath}
\usepackage[dvips,final]{graphicx}
\usepackage{url}
\usepackage{textcomp}
\usepackage{graphicx}
\usepackage{bm}
\usepackage{dcolumn}
\usepackage[usenames]{color}
\usepackage{scrextend}
\usepackage{amssymb}
\usepackage{amsmath}
\usepackage{bm}

\usepackage{slashed}
\usepackage{amssymb,amsmath}
\usepackage{lipsum}
\usepackage{mathtools}
\usepackage[dvips,final]{graphicx}
\usepackage{url}
\usepackage[font=small,skip=5pt]{caption}
\usepackage{ragged2e}
\usepackage{etoolbox}
\makeatletter
\patchcmd{\@makecaption}{\@ifdim{\wd\@tempboxa >\hsize}}{\@firstoftwo}{}{}
\makeatother
\usepackage{subcaption}
\usepackage{textcomp}
\usepackage{graphicx}
\usepackage{bm}
\usepackage{dcolumn}
\usepackage[usenames]{color}
\usepackage{amssymb}
\usepackage{amsmath}
\usepackage{amsthm}
\usepackage{amsopn}
\usepackage{titlesec}
\usepackage{xcolor}
\usepackage{axodraw2}
\usepackage{cleveref}

\def \k{\mathbf{ k}}
\def \p{\mathbf{ p}}
\def \B{\mathbf{ B}}
\def \E{\mathbf{ E}}
\def \x{\mathbf{ x}}
\def \r{\mathbf{r}}

\def\comment#1{}

\def\beq{\begin{equation}}
\def\eeq{\end{equation}}
\def\bea{\begin{eqnarray}}
\def\eea{\end{eqnarray}}

\usepackage{ulem}
\usepackage{cancel}
\usepackage{color}
\usepackage{epsfig}
\input epsf.tex
\usepackage{bm}
\usepackage{amsthm}
\usepackage{braket}
\usepackage[title]{appendix}
\usepackage{amsopn}

\def\comment#1{}

\def\mean#1{\left\langle#1\right\rangle}
\def\beq{\begin{equation}}
\def\eeq{\end{equation}}
\def\bea{\begin{eqnarray}}
\def\eea{\end{eqnarray}}

\makeatletter
\usepackage{ulem}
\usepackage{cancel}
\usepackage{color}
\def \l{\left}
\def \r{\right}
\makeatother

\begin{document}
	
	\title{\Large Probing Virtual ALPs by Precision Phase Measurements: Time-Varying Magnetic Field Background }

	\author{Mohammad Sharifian}
\email[]{mohammadsharifian@ph.iut.ac.ir}
\affiliation{Department of Physics, Isfahan University of Technology, Isfahan 84156-83111, Iran}
\affiliation{ICRANet-Isfahan, Isfahan University of Technology, 84156-83111, Iran}
	
\author{Moslem Zarei}
\email[]{m.zarei@iut.ac.ir}
\affiliation{Department of Physics, Isfahan University of Technology, Isfahan 84156-83111, Iran}
\affiliation{ICRANet-Isfahan, Isfahan University of Technology, 84156-83111, Iran}
\affiliation{Dipartimento di Fisica e Astronomia “Galileo Galilei” Universita` di Padova, 35131 Padova, Italy}
\affiliation{INFN, Sezione di Padova, 35131 Padova, Italy}

\author{Mehdi Abdi}
\email[]{mehabdi@iut.ac.ir}
\affiliation{Department of Physics, Isfahan University of Technology, Isfahan 84156-83111, Iran}

\author{ Marco Peloso}
\email[]{marco.peloso@pd.infn.it}
\affiliation{Dipartimento di Fisica e Astronomia “Galileo Galilei” Universita` di Padova, 35131 Padova, Italy}
\affiliation{INFN, Sezione di Padova, 35131 Padova, Italy}

\author{Sabino Matarrese,}
\email[]{sabino.matarrese@pd.infn.it}
\affiliation{Dipartimento di Fisica e Astronomia “Galileo Galilei” Universita` di Padova, 35131 Padova, Italy}
\affiliation{INFN, Sezione di Padova, 35131 Padova, Italy}
\affiliation{INAF - Osservatorio Astronomico di Padova, I-35122 Padova, Italy}
\affiliation{Gran Sasso Science Institute, I-67100 L'Aquila, Italy}

\date{\today}
	
\begin{abstract}
We propose an experimental scheme for detecting the effects of off-shell axion-like particles (ALPs) through optical cavities. In this proposed experiment, linearly polarized photons are pumped into an optical cavity where an external time-dependent magnetic field is present. The magnetic field mediates an interaction between the cavity photons and ALPs giving rise to a modification in the phase of the cavity photons. The time-dependent nature of the external magnetic field prompts a novel amplification effect which significantly enhances this phase modification. A detection scheme is then proposed to identify such axion-induced phase shifts. We find that the phase modification is considerably sensitive to the photon-ALPs coupling constants $g_{a\gamma\gamma}$ for the range of ALPs mass $3.1\:\mu\textrm{eV}\leqslant m_a \leqslant 44.4\:\mu\textrm{eV}$.
\end{abstract}
	
\maketitle

\section{Introduction}
Axion Like Particles (ALPs) appear in several extensions of the Standard Model (SM) and are prominent candidates for Dark Matter (DM) \cite{Abbott:1982af,Dine:1982ah,Preskill:1982cy}. They can have a multitude of different coupling to the SM particles such as electrons and photons. The particular ALP predicted by the Peccei-Quinn mechanism \cite{Peccei:1977hh,Peccei:1977ur,Weinberg:1977ma,Wilczek:1987mv} to dynamically solve the strong CP problem is called the QCD axion. The phenomenology of axions and ALPs is determined by their low mass and weak interactions with the environment. Such weak-coupling or masses below eV make up a well-motivated part of the parameter space but they make the detection of ALPs very challenging with current experimental setups. Nevertheless, the early attempts towards detection of ALPs trace back to 1980s, see e.g. \cite{Sikivie:1983ip,Sikivie:1985yu,Raffelt:1987im}. In particular, the interaction of ALPs with the photon has motivated experiments such as helioscope \cite{Lazarus:1992ry} and light-shining-through-walls experiments \cite{Chou:2007zzc,Robilliard:2007bq,Ehret:2010mh}, (see e.g. Refs.~\cite{Irastorza:2018dyq,Sikivie:2020zpn} for a review). 

The photon-ALP interaction causes a difference in phase velocity between left and right-handed circularly polarized light. In some recent proposals, it has been suggested to use high-finesse Fabry-Perot cavities to accumulate and measure the resulting phase difference \cite{Melissinos:2008vn,DeRocco:2018jwe,Obata:2018vvr,Liu:2018icu,Geraci:2018fax}. 
These cavity experiments are predicted to be sensitive to the ALP-photon coupling in the mass range around $10^{-14}\leqslant m_a  \leqslant 10^{-9}$ eV. 
ALPs can also source cosmic microwave background (CMB) birefringence \cite{Harari:1992ea,Pospelov:2008gg,Finelli:2008jv,Sigl:2018fba,Fedderke:2019ajk}. 
The interaction of CMB photons with ALPs induces a rotation in angle of the plane of CMB linear polarization.

In a recent work, some of us proposed a scheme to detect virtual axion-like particles (ALPs) via a spatially non-uniform magnetic field $\mathbf{ B}(\mathbf{ x})$~\cite{Zarei:2019sva}. We showed that this interaction leads to an enhanced birefringence phenomenon, which could be detected as a phase difference between the cavity polarization modes. The inhomogeneity of the external magnetic field caused a momentum transfer to the vertex of the interaction. 
This process was well described by an effective interaction Hamiltonian that was defined in terms of the second-order S-matrix operator \cite{Kosowsky:1994cy,Bavarsad:2009hm,Bartolo:2018igk,Bartolo:2019eac,Bartolo:2020htk}. 
We chose a spatially periodic magnetic field $\mathbf{ B}(\mathbf{ x})\propto \mathbf{ B}_0\cos(x/\ell)$ along the cavity axis ($x$-direction) with the wavelength $\ell$.  In obtaining the effective Hamiltonian describing this process, we integrated over the volume of the cavity.  Due to the presence of $\cos(x/\ell)$, integrating over $x$ gave rise to a $\rm sinc$-like profile. Therefore, in the final expression for the effective Hamiltonian we are left with an integral over $k_x$, with the integrand that was given by multiplying the propagator of the ALP and the profile function. 
The profile function developed two narrow peaks around $k_x\sim p_x \pm 1/\ell$ where $k_x$ ($p_x$) was the ALP (photon) momentum. It was shown that adjusting the photon momentum entering the cavity and the wavelength $\ell$ around these peaks amplified the phase difference significantly, such that it could be potentially measured by the cavity experiment scheme.

In this paper, we instead consider photons in an optical cavity interacting with a time-dependent external magnetic field $\mathbf{B}(t)$ mediated by an off-shell ALP.
Due to the time dependence of the external magnetic field, energy is transferred to the interaction vertex. 
Here, a harmonic magnetic field $\mathbf{ B}(t)\propto \mathbf{ B}_0 \cos(\omega_B t)$ with angular frequency $\omega_B$ is considered. Due to the presence of this oscillating magnetic field, a profile function depending on the frequency of the magnetic field and the energy of the input photons is developed in the interaction Hamiltonian.
We shall show that in addition to the Breit-Wigner enhancement \cite{Breit:1936zzb}, the profile also induces a novel amplification effect.
We provide an analytical expression for the effect of the resonance, and corroborate it with the numerical results for the impact of the resonance on the phase. The ALP mass and the frequency of the alternating magnetic field (AMF) determine the resonance frequency of photons. Therefore, one can adjust the photon frequency to have the greatest amount of interaction in order to probe each ALP mass while taking the applicable magnetic field frequency into account. Moreover, the resonance effect occurs when an AMF is combined with a constant magnetic field as well as when an AMF is applied alone. This feature makes it easier to test the theory in a real experiment.

We propose a detection scheme which benefits from a noise cancellation mechanism where the phase destroying effects such as thermal vibrations of the cavity walls, as well as the pump laser phase noise are eliminated.
For this, a Fabry-Perot cavity inside a time-varying uniform magnetic field can be employed. The cavity is aligned such that the  direction of propagation of the photons inside the cavity is perpendicular to $\textbf{B}$.
In the regime in which the electric field induced by the time-varying magnetic field is negligible, only the cavity polarizations component  aligned to the magnetic field is affected by the ALP, while the polarization mode which is perpendicular to the magnetic field remains unaffected.
Hence, by equally pumping both cavity polarization modes the axion-induced phase difference accumulated at the outgoing cavity field is detected via a balanced homodyne measurement where both output modes are combined through a 50:50 beam-splitter and  observed by two photodetectors~\cite{Gardiner}. In order to explore various ALP masses, one can change the laser frequency depending on the resonance condition of the profile function, with a constant frequency for the magnetic field. Additionally, when there is a faint data for the value of ALP mass by other experiments or theories, extending the measurement time of our scheme can improve the sensitivity to reach the desired value of coupling constant.

The paper is organized as follows: In Section II we provide a detailed calculation of the Hamiltonian describing the interaction of photons with a time-dependent magnetic field mediated by ALP. In Section III we deal with the cavity detection scheme and calculate the phase difference between cavity polarization modes. We will also present an analytical study of the new enhancement effect in this section.
Section IV is devoted to discussing the sensitivity of our cavity experiment that results in the exclusion region in the ALPs parameters. Finally, Section V presents our conclusions.


\section{ Impacts of the ALP-photon scattering on the cavity polarization modes }
The scattering process of photons off external time-dependent magnetic field $\mathbf{B}(t)$ mediated by an off-shell ALP develops a phase difference between the cavity polarization modes. It turns out that the periodicity of $\mathbf{B}(t)$ amplifies this effect. In this section, we investigate this scattering phenomenon in detail.

The following Lagrangian describes the axion-photon interaction
	\bea
	\mathcal{L}_{ a\gamma\gamma}=\frac{1}{4}\,g_{ a\gamma\gamma}\phi F_{\mu\nu}\widetilde{F}^{\mu\nu}~,
	\eea
where $\phi$ is the ALP field, $F_{\mu\nu}$ is the electromagnetic field with $\widetilde{F}^{\mu\nu}=\frac{1}{2} \epsilon^{\mu\nu\rho\sigma}F_{\rho\sigma}$ its dual, and $g_{a\gamma\gamma}$ is the coupling constant. Since we are interested in an ALP-mediated process in an external magnetic field, we split the tensor $F_{\mu\nu}$ into a homogeneous background $\bar{F}_{\mu\nu}$ and a fluctuating quantum part $f_{\mu\nu}=\partial_{\mu}A_{\nu}-\partial_{\nu}A_{\mu}$ with $A_{\mu}$ denoting the photon field.
Therefore, by expanding the interaction term about the background and neglecting the fully static and fully fluctuating terms, one arrives at the following Lagrangian
\bea
\mathcal{L}_{a\gamma\gamma}=\frac{1}{2}g_{a\gamma\gamma}\phi \,\epsilon^{\mu\nu\rho\sigma} \bar{F}_{\mu\nu}\partial_{\rho}A_{\sigma}~.
\eea

\begin{figure}
	\hspace*{-2cm}
	\centering 
	\includegraphics[width=0.5\columnwidth]{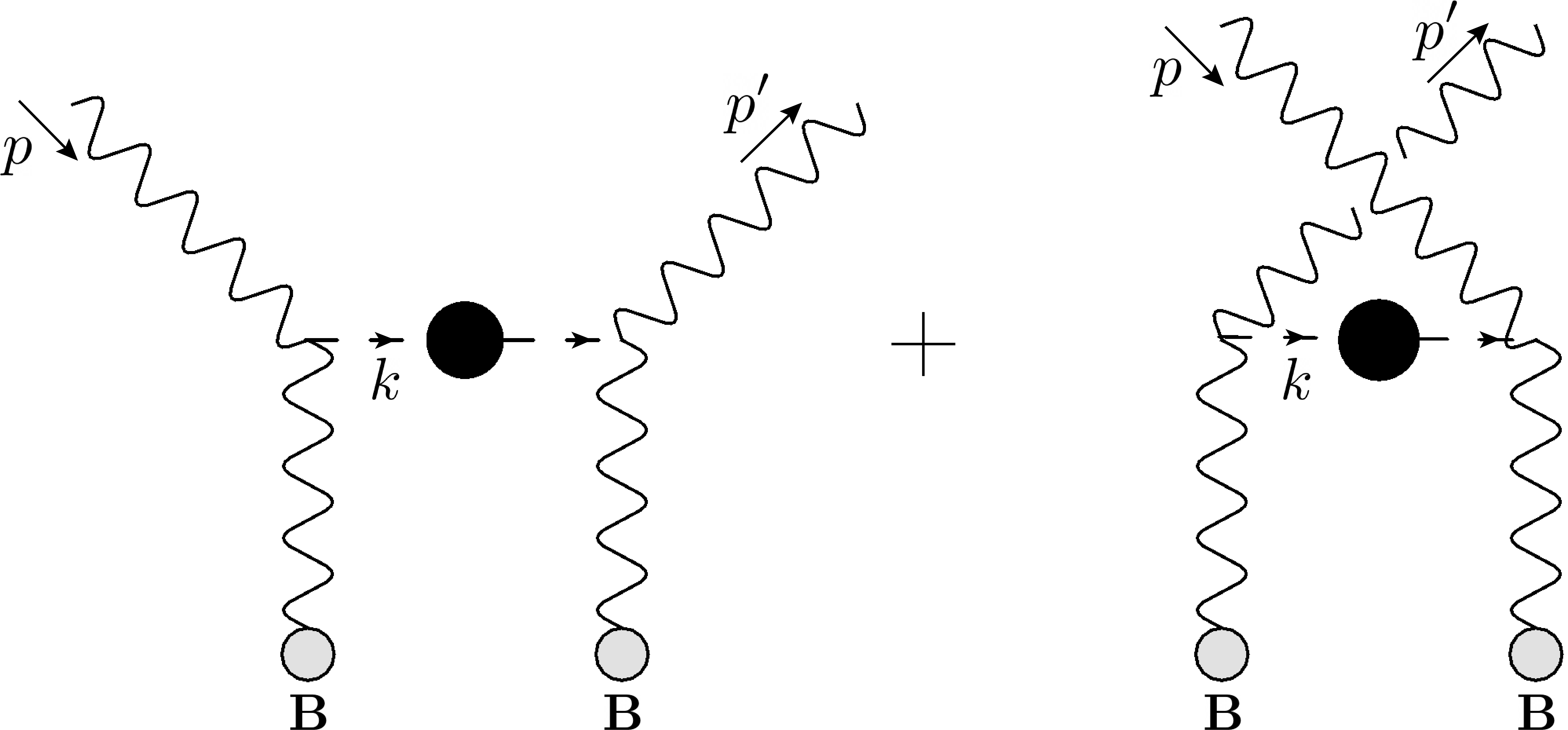}
	\caption{\small Two Feynman diagrams corresponding to the scattering process.}  \label{Feyn}
\end{figure}
This interaction Lagrangian allows us to investigate the interaction of a propagating polarized photon with a time-dependent background magnetic field through the intermediate ALP field.
Fig. \eqref{Feyn} illustrates the two Feynman diagrams associated with such process. The effective interaction Hamiltonian is found through the S-matrix operator associated with this process \cite{Kosowsky:1994cy}
\bea
	S^{(2)}(\gamma_s\rightarrow\gamma_{s'})=-i\int dt \,H(t)~,
\eea
where $H(t)$ describing the process shown in Fig.~\eqref{Feyn} is given by
		\bea  \label{Hint1}
		H(t)&=&\frac{g^2_{ a\gamma\gamma}}{8}\int d^4x' d^3x   \epsilon^{\mu\nu\rho\sigma}\epsilon^{\mu'\nu'\rho'\sigma'} \bar{F}_{\rho\sigma} (x)\bar{F}_{\rho'\sigma'} (x')D_F(x-x')
		\left [\partial_{\mu}A^{-}_{\nu}(x) \partial_{\mu'}A^{+}_{\nu'}(x') 
	 \right.	\nonumber \\  && \left. ~~~~~~~~~~~~+~
		 \partial_{\mu'}A^{-}_{\nu'}(x') \partial_{\mu}A^{+}_{\nu}(x)\right ]~.
		\eea
Here, $D_F(x-x')$ is the Breit-Wigner Feynman propagator for the ALP, and $A^{+}_{\mu}$ ($A^{-}_{\mu}$) is the electromagnetic field linear in the absorption (creation) operators of the photons. 
It is worth mentioning that due to the time-dependence of the external magnetic field, an electric field is induced as implied by Faraday's law $\bm{\nabla} \times \mathbf{ E}=-\frac{\partial}{\partial t}\,\mathbf{ B}$.
Therefore, one must keep both magnetic field $B^k(t)=-\epsilon^{k i j}\bar{F}_{ij}/2$ and electric field $E_i=F_{0i}$ contributions in the interaction Hamiltonian \eqref{Hint1}. However, as we shall discuss below, the induced electric field can be made negligible by a proper choice of setup parameters.

We consider a one-dimensional cavity. This is effectively realized by taking one direction of the quantization boundaries, say $L_x$, to be much smaller than the other two, $L_y$ and $L_z$. The cavity is placed within a time dependent background magnetic field aligned along the $z$-direction. Now, we estimate the induced electric field inside the cavity. Starting from Faraday's law,
	\begin{equation} \label{FLaw}
	\nabla\times{\E}= - \frac{\partial \B}{\partial t}~,
	\end{equation}
For an oscillating magnetic field of the angular frequency $\omega_B$ and the amplitude $B_0$ we can also write $\frac{\partial }{\partial t}\B \sim B_0\,\omega_B$. According to Eq. \eqref{FLaw}, one roughly has  $E_x \sim B_0\,\omega_BL_y$ and $E_y \sim B_0\,\omega_BL_x$, where $E_x$ ($E_y$) is the amplitude of the induced electric field and $L_x$ ($L_y$) is the cavity length along $x$ ($y$) axis. When $L_x$ is taken as the cavity's smallest length scale, $E_x$ can be higher than $E_y$. The induced electric field  along $x$ or $y$ axis can be calculated as
	\beq
	\frac{E_{x,y}}{B_0} \sim  4\times 10^{-5}\frac{\omega_B}{250\,\textrm{kHz}} \frac{L_{x,y}}{10\,\textrm{cm}}~,
	\eeq
which indicates that the induced electric field will be subdominant within the cavity length ranges that we will consider in our scheme. Otherwise, for much larger cavity lengths the electric field becomes significant, and it must be involved in the interaction Hamiltonian. The effect of $E_y$ is a phase deviation on the z-polarized component of the photon but it is negligible because the E-field is much smaller than the applied B-field. In this work, we only take parameter ranges that the electric field can be neglected.

To evaluate the Hamiltonian \eqref{Hint1} in this background magnetic field, we quantize the photon field $A$ in the coulomb gauge,  $A^0 = 0$. The photon field quantization in a cavity formed by perfectly conducting walls has been discussed in~\cite{kakazu:1994cy}. Here we only assume that the smaller length $L_x$ is bounded by two conducting walls, while we treat the other dimensions as unbounded (consistently with the $L_y = L_z \gg L_x$ setup). The boundary conditions on $\textbf{A}(x)$ are thus applied so that the tangential component of the electric field associated with $\textbf{A}(x)$ and the normal component of the magnetic field associated with $\textbf{A}(x)$ must vanish at the cavity boundaries in this direction. By imposing these boundary conditions, we arrive at
\begin{eqnarray}      \label{A+0}
	\textbf{A}(x)=\sum_{s}\sum_{\textbf{p}}  \frac{1}{\sqrt{\omega_{\p} V}}
		\left[a_s(\p) \mathbf{ u}_s(\p)+a^{\dagger}_s(\p) \mathbf{ u}^{*}_s(\p)  \right]~,
\end{eqnarray} 
where the mode functions are given by
\begin{eqnarray}
	\mathbf{ u}_s(\p)=\left(\cos(p_{x} x)\varepsilon_{s}^x(\p),\,i\sin(p_{x} x)\varepsilon_{s}^y(\p),\,i\sin(p_{x} x)\varepsilon_{s}^z(\p) \right)\,e^{i p_y y}\,e^{i p_z z}\,e^{-i\omega_{\textbf{p}}t}~,
\end{eqnarray}
where $\bm{\varepsilon}_s$ is the photon polarization, $\p=\left(p_{x},p_{y},p_{z}\right)=\left(\frac{\pi l_1}{L_x},\frac{2\pi l_2}{L_y},\frac{2\pi l_3}{L_z}\right)$ with $l_2,l_3=0,\pm1,\pm2,  \cdot\cdot\cdot$ and $l_1=0, 1, 2, \cdot\cdot\cdot$. In Eq. \eqref{A+0} we also have annihilation and creation operators of the photons with polarization $s$ and momentum $\p$, that satisfy
\beq \label{modefunction}
\l[a_s(\p),a^{\dagger}_{s'}(\p')\r]=\delta_{ss'}\delta_{\p,\p'}~.
\eeq
Finally, $\omega_{\textbf{p}}=|\textbf{p}|$ is the photon angular frequency, while $V$ is the cavity volume.

Inserting the expression \eqref{A+0} in Eq.~\eqref{Hint1}, and assuming a background magnetic field, results in two scalar products between the magnetic field and the polarization vector of $\textbf{A}(x)$. Since the magnetic field is oriented along the $z$-direction, we can effectively replace \eqref{A+0} with the simpler \begin{eqnarray}    \label{A+1}
			\textbf{A}(x)=\textbf{A}^+(x) + \textbf{A}^-(x)=\sum_{s}\sum_{\textbf{p}}\frac{1}{\sqrt{\omega_{\textbf{p}} V}}\left[ i\sin(p_{x} x)\,a_s(\p)\,e^{i p_y y}e^{i p_z z}e^{-i\omega_{\textbf{p}}t}\,\varepsilon_s^z(\p)+ \text{h.c.}  \right]~.
\end{eqnarray}
Eq.~\eqref{Hint1} then evaluates to 
\bea \label{H}
H(t)&=&\frac{g^2_{a\gamma\gamma}}{2 V}
\,\sum_{s,s'}\sum_{\p ,\p'}\sqrt{\omega_\mathbf{p}\omega_\mathbf{\p'}}
\int dt'dx'd^2 \x'_\bot \,dx\, d^2 \x_\bot\,\frac{d^4k}{(2\pi)^4}\,D_F(k)
\left[\bm{\varepsilon}_{s}(\p)\cdot\textbf{B}(t)\right]\left[\bm{\varepsilon}_{s'}^*(\p')\cdot\textbf{B}(t')\right]
\nonumber \\ &\times &
a_{s'}^{\dagger}(\p')a_{s}(\p) \left[\sin(p'_{x} x)\sin(p_{x} x')
e^{-i\k\cdot \x' }
e^{-i \textbf{p}'_{\bot}\cdot   \x_\bot}
e^{i\k\cdot \x }
e^{i \textbf{p}_\bot\cdot   \x'_\bot}
e^{-i(-k^0+\omega_{\mathbf{p}})t'}
e^{-i(k^0-\omega_{\mathbf{p}'})t}
\right.\nonumber\\
&+& \left. \sin(p'_{x} x')\sin(p_{x} x)
e^{-i\k\cdot \x' }
e^{-i \textbf{p}'_{\bot}\cdot   \x'_\bot}
e^{i\k\cdot \x }
e^{i \textbf{p}_\bot\cdot   \x_\bot}
e^{-i(-k^0-\omega_{\p'})t'}
e^{-i(k^0+\omega_{\p})t}\right]~,
\eea
where $D_F(k)$ is the Fourier transform of $D_F(x)$, which is defined by (see Appendix \ref{appendix:propagator})
\bea \label{Dx1}
D_F(x-x')=\int \frac{d^4k}{(2\pi)^4}\frac{i}{k^2-m_a^2+ik^0\Gamma_B(k^0)}e^{-ik\cdot (x-x')}~,
\eea
in which $m_a$ is the ALP physical mass, and $\Gamma_B(k^0)$ is the ALP decay rate in the presence of a time-dependent magnetic field and has been derived in Appendix \ref{appendix:decayrate}.

	
\subsection{ An alternating magnetic field}
\label{section:AMF}

The central quantity for describing the process described above is the interaction Hamiltonian $H(t)$, defined from the S-matrix. Therefore, it is crucial to construct well-separated in and out states with the specified content in the far past and future. The in and out states must be time-independent. This cannot be realized in the presence of a time-dependent background field. We consider a setup where the interaction between photon and background field is turned on (off) at $-\frac{1}{2}\tau_B$ ($+\frac{1}{2}\tau_B$).
Therefore, the time-dependent magnetic field is considered as the form
\begin{eqnarray}\label{bg}
		\mathbf{B}(t)= \mathbf{B}_{0}\,\cos{\omega_B t}\,\Pi\l(\frac{t}{\tau_B}\r)~,\;\;\;\;\;\;\;\;\,\B_0=B_0 (0,0,1)~\,
	\end{eqnarray}
where $\Pi(t/\tau_B)$ is a rectangular function centered at $t=0$ that models the time duration $\tau_B$ in which the magnetic field is applied. We will also compare our results with a smoother approximation of $\Pi(t/\tau_B)$ that is closer to the real conditions. 

In Appendix \ref{appendix:interaction}, we have a detailed derivation of the interaction Hamiltonian for the above magnetic field based on Eq.~\eqref{H}. The Hamiltonian then reads
  \bea \label{Hint1-3}
		H(t)=\sum_{s,s'}\sum_{\p}\mathcal{F}(t)\,\l[\bm{\varepsilon}_{s'}^*(\p)\cdot\hat{\mathbf{b}}\r]
		\l[\bm{\varepsilon}_{s}(\p)\cdot\hat{\mathbf{b}}\r]a_{s'}^{\dag}(\p) a_s(\p)~,
		\eea 
in which $\hat{\mathbf{b}}$ is the direction of magnetic field and $\mathcal{F}(t)$ is
		\bea \label{Ft1}
		\mathcal{F}(t)&=&G_a\,\cos(\omega_B t)\,\Pi\l(\frac{t}{\tau_B}\r)\,\omega_{\mathbf{p}}\,\int_{-\infty}^\infty dk^0
		\frac{1}{{k^0}^2-\tilde{\omega}_{\mathbf{p}}^2+ik^0\Gamma_B(k^0)}\l[e^{-it(k^0-\omega_{\mathbf{ p}})}
		P_1(k^0)
		\right. \nonumber \\ &+& \left. e^{-it(k^0+\omega_{\mathbf{ p}})}
		P_2(k^0)\r]~,
		\eea
where $G_a= \frac{g^2_{ a\gamma\gamma}B_0^2}{32\pi} $, $\tilde{\omega}^2_{\mathbf{p}}=\omega^2_{\mathbf{p}}+m_a^2$, and the two profile functions $P_1(k^0)$ and $P_2(k^0)$ are defined as
\bea \label{P1}
P_1(k^0)=\int_{-\infty}^{\infty}  dt' \cos(\omega_Bt')\,\Pi\l(\frac{t'}{\tau_B}\r)e^{-i(-k^0+\omega_{\p})t'}=\frac{\sin{\left(\Delta_1\tau_B/2\right)}}{\Delta_1}+\frac{\sin{\left( \Delta_2 \tau_B/2\right)}}{\Delta_2}~,
\eea
and
\bea  \label{P2}
P_2(k^0)=\int_{-\infty}^{\infty}  dt' \cos(\omega_B t')\,\Pi\l(\frac{t'}{\tau_B}\r)e^{-i(-k^0-\omega_{\p})t'}=
\frac{\sin{\left(\Delta_3\tau_B/2\right)}}{\Delta_3}+\frac{\sin{\left( \Delta_4 \tau_B/2\right)}}{\Delta_4}~,
\eea
with $\Delta_1=k^0+\omega_B-\omega_{\mathbf{p}}$, $\Delta_2=k^0-\omega_B-\omega_{\mathbf{p}}$
, $\Delta_3=k^0+\omega_B+\omega_{\mathbf{p}}$, and $\Delta_4=k^0-\omega_B+\omega_{\mathbf{p}}$.

The function $\mathcal{F}(t)$ entering in the Hamiltonian \eqref{Hint1-3} is a time-dependent coupling that as we show below causes an enhancement effect in the phase difference between two polarizations of the photons inside the cavity.

\begin{figure}
\centering   
\includegraphics[width=.7\textwidth]{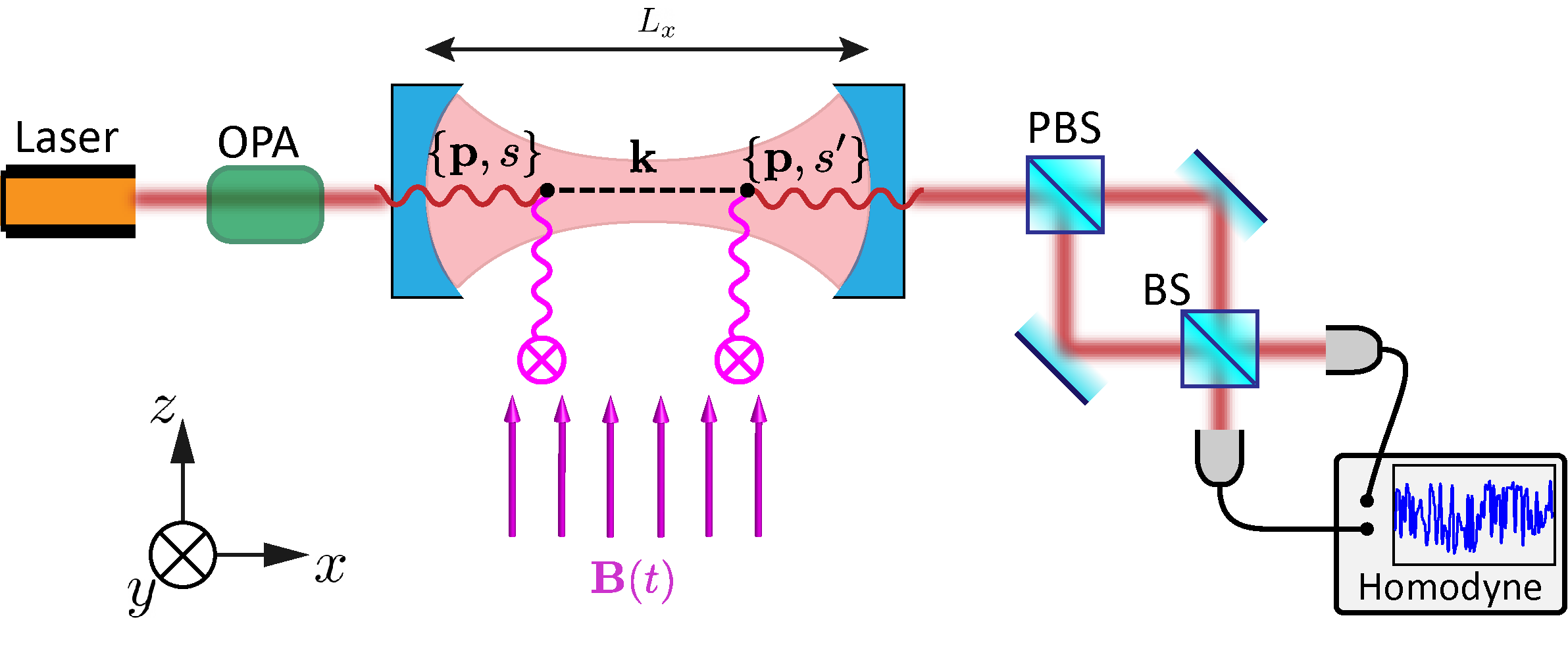}
\caption{\small  The schematic diagram of setup proposed in this paper: An optical cavity of length $L_x$ is pumped by coherent or squeezed light which is amplified by an Optical Parametric Amplifier (OPA). A background magnetic field, $\mathbf{B}(t)$, is applied perpendicularly to the cavity axis, mediating the laser interaction with ALPs. The existence of an alternating magnetic field causes an enhancement on the interaction of the photon (with momentum $\p$ and polarization indices $s$ \& $s'$) with the virtual ALPs (with momentum $\k$). The Polarized Beam Splitter (PBS) and the Beam Splitter (BS) split the two linearly polarized beams into two separated beams. The axion-induced dispersion introduces a phase deviation which is measured through a homodyne detection.}
\label{scheme}
\end{figure}

\subsection{Combination of a constant and an alternating magnetic field}\label{section:CAMF}
Since generating high magnitudes of alternating magnetic fields is not feasible with the current technology, one can consider an alternating magnetic field which is modulated on top of a constant magnetic field offset as
\begin{eqnarray}
	 \mathbf{B}(t)= \left(B_1+B_{0}\,\cos{\omega_B t}\,\Pi\l(\frac{t}{\tau_B}\r)\right)\,\hat{\mathbf{b}}~,\;\;\;\;\;\;\;\;\hat{\mathbf{b}}=(0,0,1)~\,.
\end{eqnarray}
Following the same procedure as \eqref{H}-\eqref{Hint1-3} for the effective interaction Hamiltonian
\begin{eqnarray} \label{DCAC}
\mathcal{F}(t)&=&\frac{g^2_{ a\gamma\gamma}}{32\pi}\Bigg(4\pi \,B_1^2\,\omega_{\p}\,\frac{m^2+i\omega_{\p}\Gamma_B(\omega_{\bf p})}{m^4+\omega_{\p}^2\Gamma_B^2(\omega_{\p})}\nonumber\\&+&4\pi B_0B_1\cos(\omega_B t)\Pi\l(\frac{t}{\tau_B}\r)\omega_{\mathbf{p}}\cos(\omega_{\p}t)\l[\frac{e^{-it\omega_{\p}}}{-m^2+i\omega_{\p}\Gamma_B(\omega_{\p})}\r]\nonumber\\&+&B_0B_1\omega_{\mathbf{p}}\int_{-\infty}^\infty dk^0
\frac{e^{-it(k^0-\omega_{\mathbf{ p}})}
	P_1(k^0)+e^{-it(k^0+\omega_{\mathbf{ p}})}
	P_2(k^0)}{{k^0}^2-\tilde{\omega}_{\mathbf{p}}^2+ik^0\Gamma_B(k^0)}\nonumber\\&+&B_0^2\cos(\omega_B t)\Pi\l(\frac{t}{\tau_B}\r)\omega_{\mathbf{p}}\int_{-\infty}^\infty dk^0
\frac{e^{-it(k^0-\omega_{\mathbf{ p}})}
	P_1(k^0)+e^{-it(k^0+\omega_{\mathbf{ p}})}
	P_2(k^0)}{{k^0}^2-\tilde{\omega}_{\mathbf{p}}^2+ik^0\Gamma_B(k^0)}\Bigg)~.\nonumber\\
\end{eqnarray}
As it is discussed below, the signal enhancement effect stems from the terms that comprise the profile functions $P_1(k^0)$ and $P_2(k^0)$, the third and fourth lines in the above equation.


\section{Cavity detection scheme}

Our strategy for tracing polarization flips induced by the ALPs is to use high-finesse cavities and high-precision phase measurement schemes. This is schematically illustrated in Fig.~\eqref{scheme}; a Fabry-Perot optical cavity of length $L_x$ is placed in an external time-dependent magnetic field that is perpendicular to the cavity axis. 
A high-frequency magnetic field, that satisfies the resonance conditions discussed below, influences the cavity fields in a birefringence fashion.

\begin{figure*}[t]
	\includegraphics[width=0.4\columnwidth]{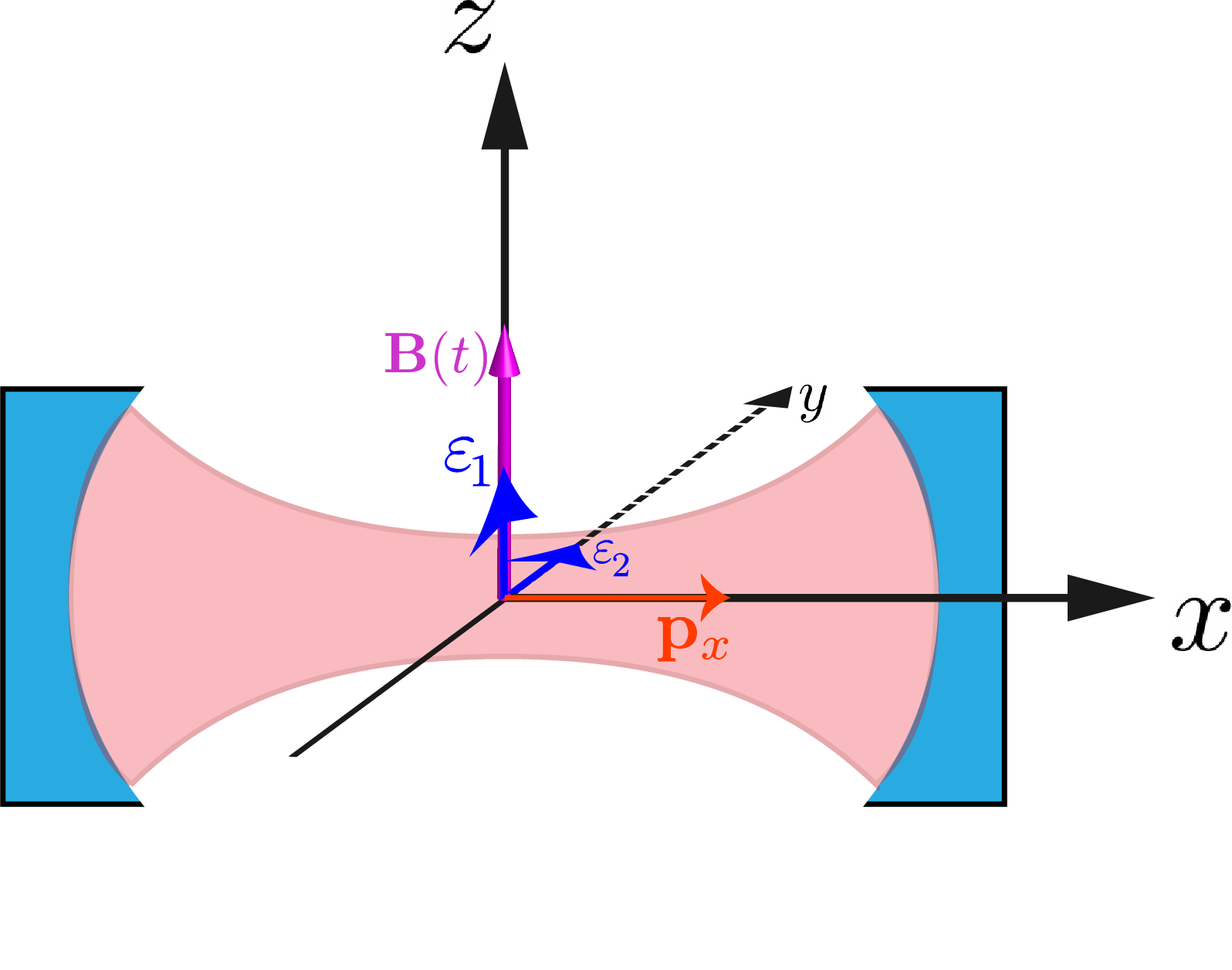}
	\includegraphics[width=0.35\columnwidth]{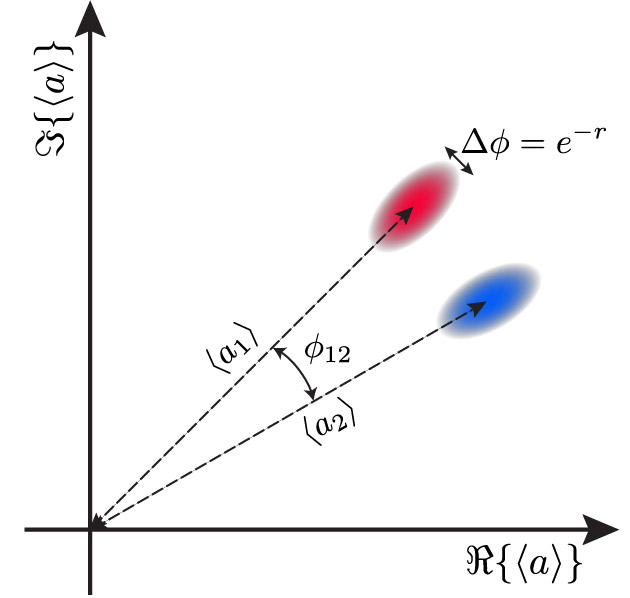}
	\caption{The polarization vectors are chosen to be in the $y$ and $z$ directions (left). The phase space diagram of the two polarization modes (right).}
	\label{polvec}
\end{figure*}

\begin{figure*}
 	\centering


 \captionsetup[subfigure]{oneside,margin={.5cm,0cm}}
 \begin{subfigure}[t]{.485\textwidth}
	\hspace*{-.3cm}
	\includegraphics[width=\textwidth]{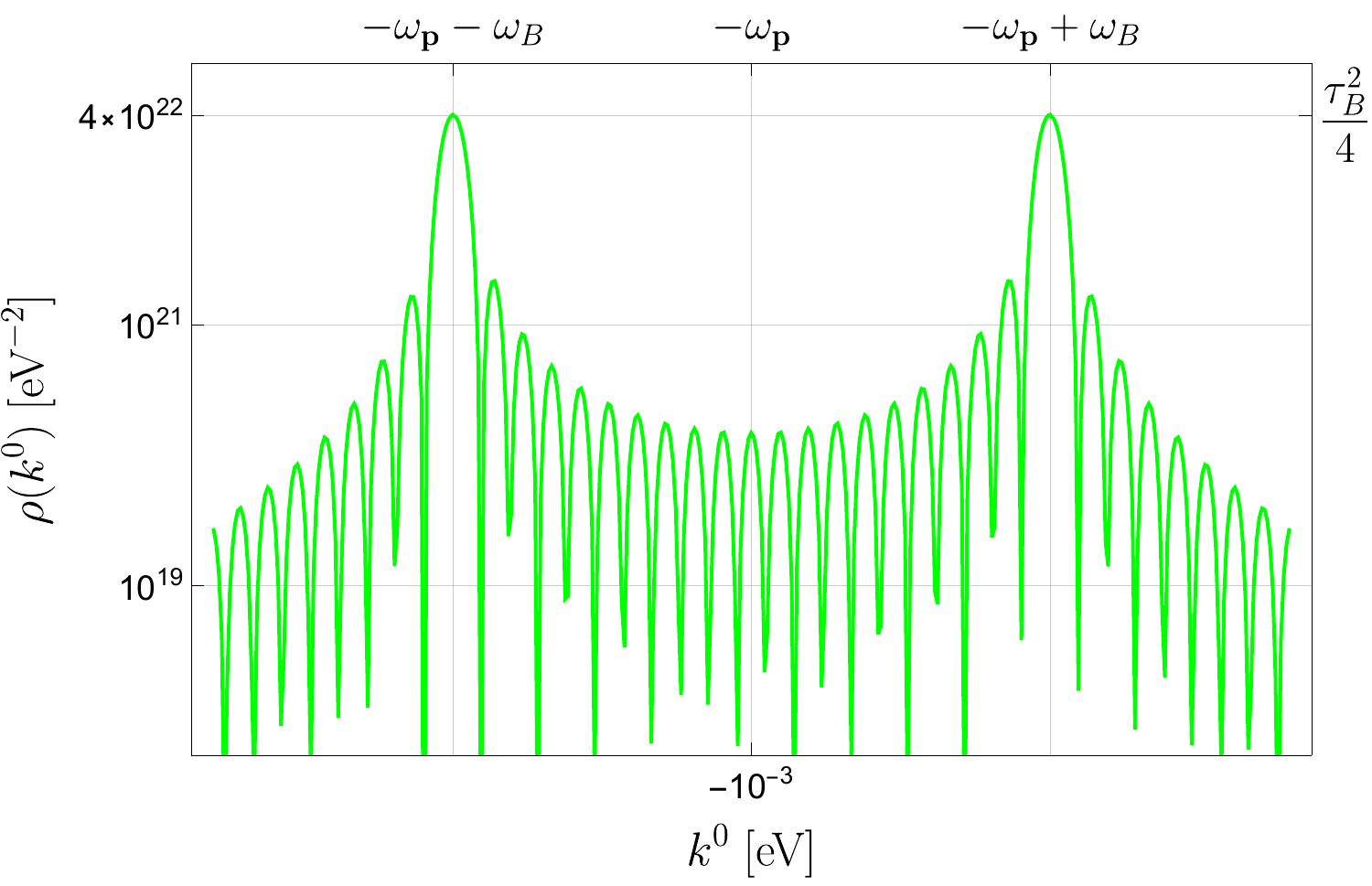}
	\caption{}
	\label{profile:fig1}
\end{subfigure}
\quad
\captionsetup[subfigure]{oneside,margin={.7cm,0cm}}
\begin{subfigure}[t]{.485\textwidth}
	\includegraphics[width=\textwidth]{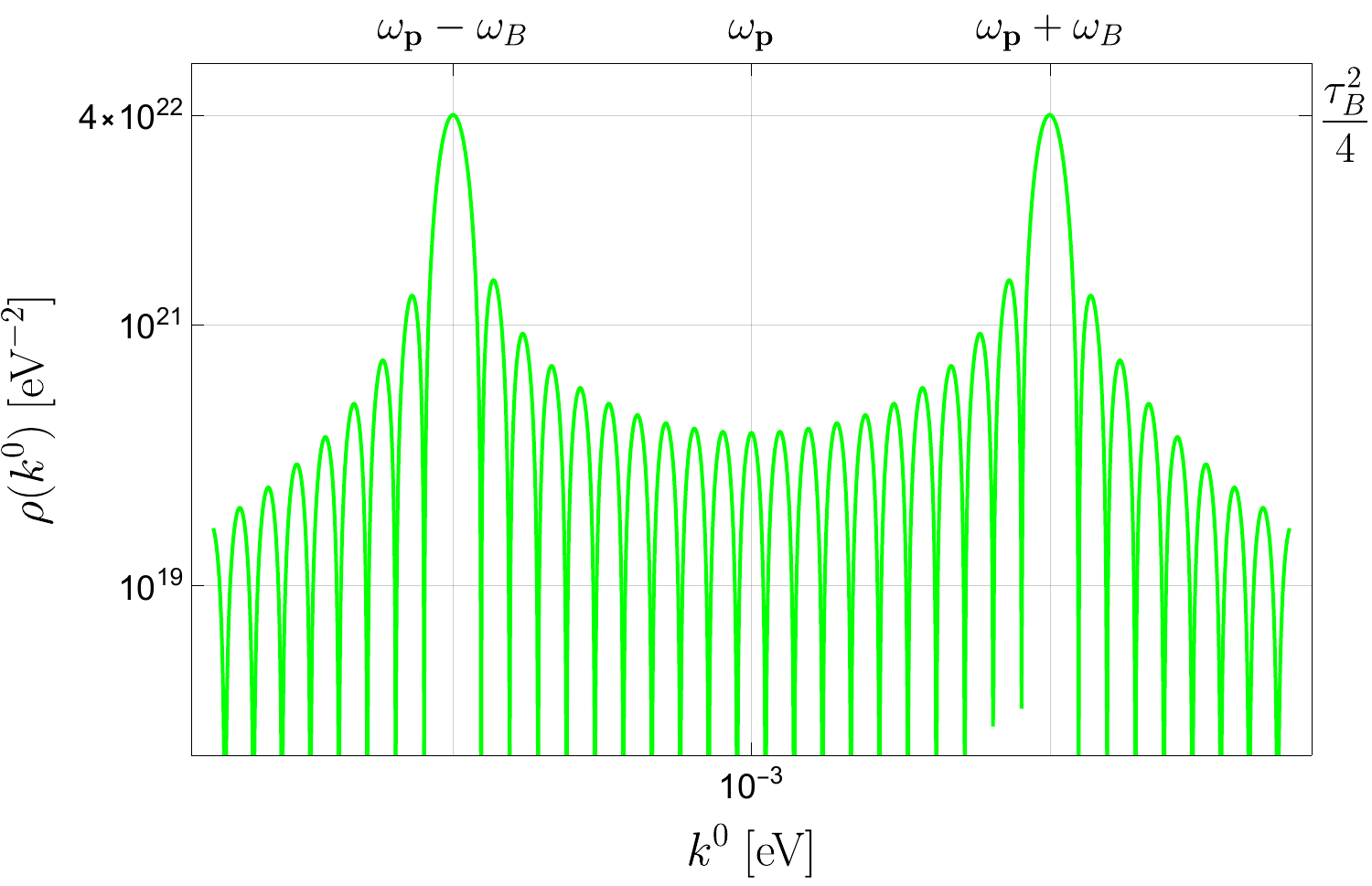}
	\caption{}
	\label{profile:fig2}
\end{subfigure}
	\\
 \captionsetup[subfigure]{oneside,margin={1.1cm,0cm}}
 	\begin{subfigure}[t]{.48\textwidth}
 		\hspace*{-.3cm}
 		\includegraphics[width=\textwidth]{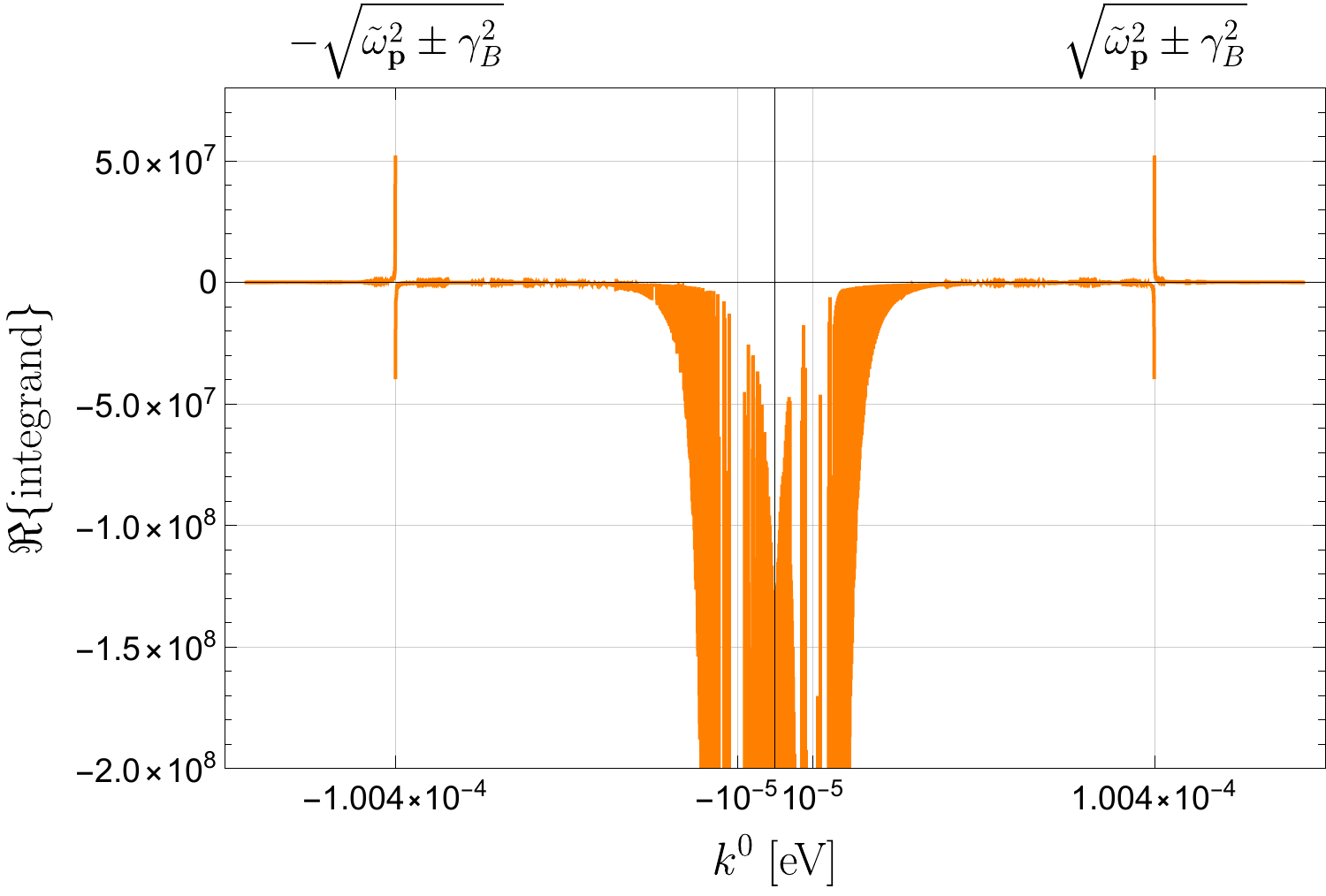}
 		\caption{}
 \label{integrand-re:fig}
 	\end{subfigure}
 \quad
  \captionsetup[subfigure]{oneside,margin={1.45cm,0cm}}
  	\begin{subfigure}[t]{.48\textwidth}
 	\includegraphics[width=\textwidth]{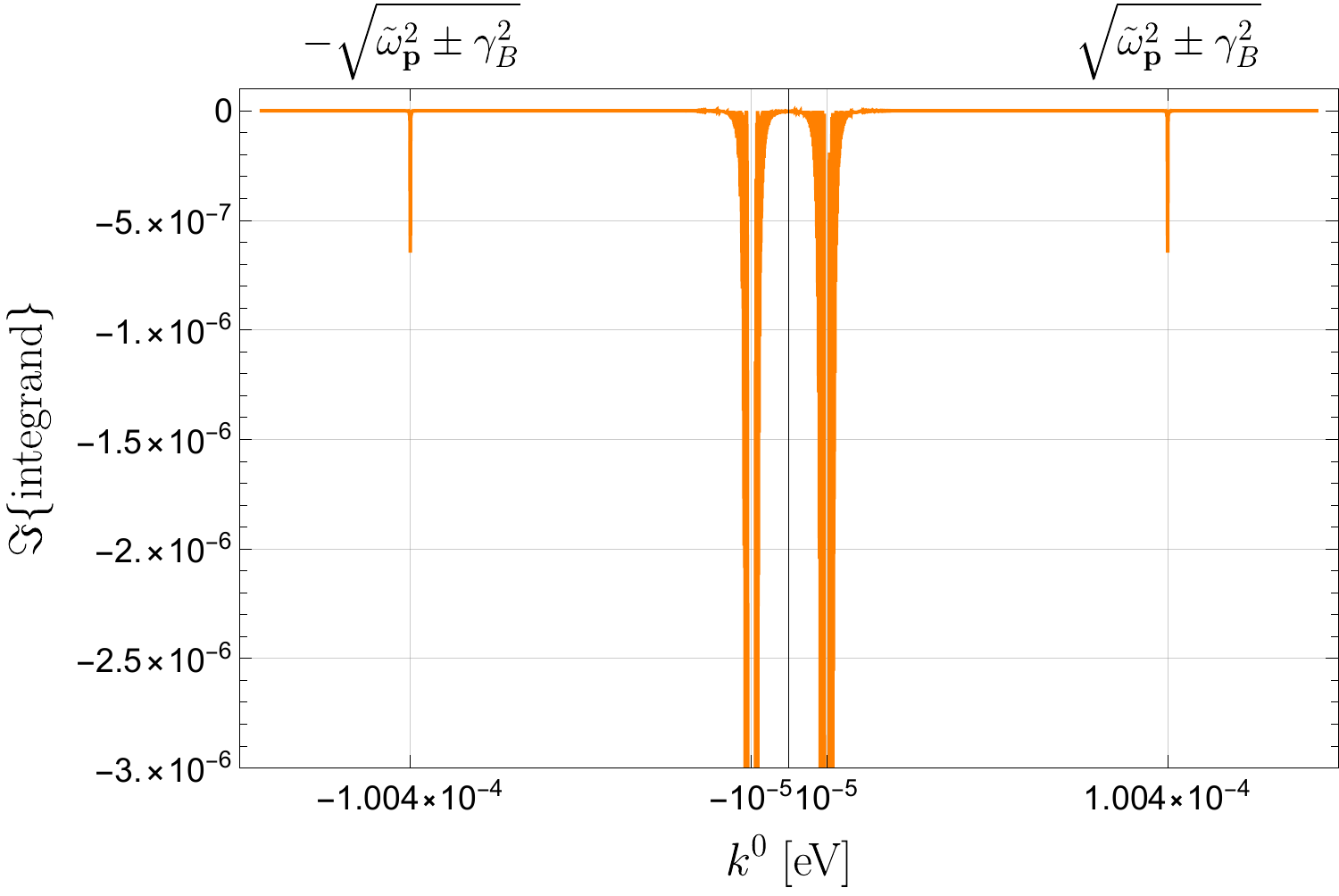}
	\caption{}
 	\label{integrand-im:fig}
 \end{subfigure}
 	\caption{ \small (a,b) The profile function for $\omega_\p=10^{-2}$\,eV and $\omega_B=250$\,kHz = $1.6\times10^{-10}$\,eV. By reducing the $\omega_B$,  the four peaks merge two by two and the resulting plot will have just two sharp peaks on $\pm \omega_{\mathbf{ p}}$. (c) The real part and (d) the imaginary part of the integrand in \eqref{intFt} for $\omega_\p=10^{-5}$\,eV, $\omega_B=100$\,kHz, $m=10^{-4}$\,eV, $g_{a\gamma\gamma}=10^{-10}$\,GeV$^{-1}$, and $\gamma_B=1.5\times10^{-25}\;\text{eV}^2$. }  \label{profile}
 \end{figure*}

The Hamiltonian of our system is given by
\beq
H_{\textrm{S}}=\sum_{s=1,2}\sum_{\p}\omega_{\p}a_s^{\dag}(\p)a_s(\p)~,
\label{Hsystem}
\eeq
describing doubly degenerate cavity modes with two orthogonal polarizations.
In the presence of an oscillating magnetic field, ALPs interact with the cavity modes depending on the angle of their polarization vector with respect to the magnetic field: A full alignment leads to the maximum interaction, while a photon polarized perpendicular to the magnetic field is unaffected.
Without loss of generality, we consider a geometry where only one of the polarization modes `feels' the ALPs through our proposed interaction scheme, and the other degenerate perpendicular polarization is employed as a phase reference, see Fig.~\eqref{polvec}. The total interaction Hamiltonian is thus written as
\beq
H_{\textrm{int}}=\sum_{\textbf{q}}\mathcal{F}(t) a^\dagger_1(\textbf{q})a_1(\textbf{q})~,
\label{Hinteraction}
\eeq
where we denoted by $\mathbf{q}$ the momentum in this expression, not to confuse it with the momentum $\p$ of the cavity photons in the experiment. We can now use this interaction Hamiltonian to study the time evolution of the cavity photons of momentum $\p$. The environmental effects are taken into account as dissipation and fluctuations in the modes following the standard input-output theory~\cite{Gardiner:1985,Gardiner,Walls}. We strat from the Langevin-Heisenberg equation for the system operators
\begin{align} \label{langevin}
	\frac{da_i(\p)}{dt}=-i[a_i(\p),H_S+H_{\text{int}}]-\kappa\, a_i(\p)+\sqrt{2\kappa}\,a_i^{\text{in}}~,
\end{align}
in which $a_i^{\textrm{in}}$ is the input field operator and $\kappa$ is the bare coupling of the system and the environment known as the cavity decay rate. This parameter determines the photon lifetime (the average time spent by the photons inside the cavity) through $\tau=1/\kappa$. The following Langevin equations for the cavity mode operators result from Eq.~\eqref{langevin},
\begin{subequations}
\begin{align}
\label{Langevin1}
\dot{a}_1(\textbf{p})=-\left[\bar{\kappa}+i\bar{\omega}_\p\right]a_1(\textbf{p}) + \sqrt{2\kappa}\,a_1^{\textrm{in}}~, \\
\dot{a}_2(\textbf{p})=-\left[\kappa+i\omega_\p \right]a_2(\textbf{p}) + \sqrt{2\kappa}\,a_2^{\textrm{in}}~,
\label{Langevin2}
\end{align}\label{Langevins}
\end{subequations} where $\bar{\kappa}$ and $\bar{\omega}_\p$ are defined as
\beq
\bar{\kappa}=\kappa-\mathcal{F}_i~,~~~~~\bar{\omega}_\p=\omega_\p+\mathcal{F}_r~,
\eeq
with $\mathcal{F}_r$ and $\mathcal{F}_i$ as the real and imaginary parts of $\mathcal{F}(t)$, respectively. Note that in the above Langevin equations, $a_i(\textbf{p})$ appears with a coefficient of mass-dimension $M$, while $a^{\text{in}}_i$
appears with a coefficient of dimension $M^\frac{1}{2}$. However, all the terms in the equation have the same dimension. This results from the convention which has been adopted in the quantum optical input-output theory. Now, a laser pulse on-resonance with one of the cavity normal frequencies $\omega_{\mathbf{p}}=n\pi c/L_x$ pumps the cavity. Here, $\pi c/L_x$ is the
free spectral range (FSR) of the cavity. The equation \eqref{Langevin1} shows that $\mathcal{F}(t)$ only affects the dynamics of the photons polarized along the magnetic field. The imaginary part of $\mathcal{F}(t)$ contributes to the cavity decay rate and potentially gives rise to the amplification of the output signal. On the other hand, the real part of $\mathcal{F}(t)$ contributes to the phase of the output signal and produces a phase difference with respect to the signal with the polarization perpendicular to the magnetic field. Eqs.~\eqref{Langevins} are solved by
\begin{subequations}
\begin{align}
\label{a1}
a_1(t) &=e^{-\int_{t_0}^{t} dt_1\left(\kappa+i\omega_{\textbf{p}}+i\mathcal{F}(t_1)\right)}\left[a_1(t_0)+\sqrt{2\bar\kappa}
\int_{t_0}^t e^{-i\omega_\p t_2+\int_{t_0}^{t_2}dt_1\left(\kappa+i\omega_{\textbf{p}}+i\mathcal{F}(t_1)\right)}a_1^{\textrm{in}}(t_2)dt_2\right]~, \\
\label{a2}
a_2(t) &=e^{-\int_{t_0}^{t} dt_1\left(\kappa+i\omega_{\textbf{p}}\right)}\left[a_2(t_0)+\sqrt{2\kappa}
\int_{t_0}^t e^{-i\omega_\p t_2+\int_{t_0}^{t_2}dt_1\left(\kappa+i\omega_{\textbf{p}}\right)}a_2^{\textrm{in}}(t_2)dt_2\right]~.
\end{align}
\end{subequations}
Below, we evaluate these equations numerically. Before doing so, however, we can study these formal solutions analytically, to get a better insight into the origin of the enhancement signal effect that we will find numerically. We first turn to the time integration over $\mathcal{F}(t)$ that appears in \eqref{a1}. 
For the input fields at $t_0\rightarrow -\infty$ and the output fields at $t \rightarrow \infty$ we take integration over $t_1$ as
\beq \label{intFt}
\varphi\equiv \int_{-\infty}^{\infty} dt_1\mathcal{F}(t_1)=G_a\omega_\p 	\int_{-\infty}^\infty dk^0
		\frac{\rho(k^0)}{{k^0}^2-\tilde{\omega}_{\mathbf{p}}^2+i\gamma_B}~,
\eeq
where $\gamma_B=k^0\Gamma^{\textrm{max}}_B(k^0)$ and
\beq \label{rho}
\rho(k^0)=P_1^2(k^0)+P_2^2(k^0)~,
\eeq
is a spectral density-like function. As we will show below, the integrand of \eqref{intFt} develops extra peaks, besides those from the propagator, due to the function $\rho(k^0)$. 
The pole structure of the propagator and the analytic properties of $\rho(k^0)$ around its peaks is of great importance towards understanding the enhancement effect that occurs in the axion-induced phase. 
The function $\rho(k^0)$ is different from the standard spectral density that appears in quantum field theory.
 In the standard spectral density, there is a Dirac delta function which corresponds
to the existence of a one-particle state in the spectrum of the theory. This term is then followed by a second term, that has the structure of a continuum component. 
Here, the propagator is weighted with $\rho(k^0)$ which plays as a moderator. 
In Figs.~\eqref{profile:fig1} and \eqref{profile:fig2} we plot $\rho(k^0)$ with respect to $k^0$ for $\omega_{\textbf{p}}=10^{-2}$~eV and $\omega_{B}=250$~kHz~\cite{HofmannPat1985,CostaPat1987,TaylorPat,schwarzPat}. We recognize that there are four positive sharp peaks at 
$k^0=\omega_\p\pm \omega_B$ and $k^0=-\omega_\p\pm \omega_B$. In Figs. \eqref{integrand-re:fig} and \eqref{integrand-im:fig}, we then display, respectively, the real and the imaginary parts of the integrand in Eq. \eqref{intFt}. The left and right panels show the real and imaginary parts of the integrand, respectively. 
In the real part, there are four sharp side peaks at $k^0= \pm\sqrt{\tilde{\omega}^2_\mathbf{p} \pm \gamma^2_B }$ produced by the poles of the propagator. The integrand changes sign on each side by moving from $\sqrt{\tilde{\omega}^2_\p - \gamma^2_B}$ to $\sqrt{\tilde{\omega}^2_\p +\gamma^2_B}$. In the intermediate region close to $k^0 = 0$, the profile effects become more important.
In this region, four peaks corresponding to the peaks of the profile function $\rho(k^0)$ appear. 
The same structure is observed for the imaginary part, with the difference that the imaginary part does not change sign across the poles from the propagator.

To proceed with our analytic estimate of the solution, we first focus on the side peaks arising from the propagator. We perform an expansion of the full propagator around its poles and compute 
the integration over $k^0$ in \eqref{intFt} close to the poles as follows
\bea \label{integralk0}
		\frac{\varphi}{G_a\omega_\p}\Big\vert_{\rm from\;poles\;of\;propagator}&=& \textrm{not enhanced }
		+\frac{\rho(-\tilde{\omega}_{\p})}{-2\tilde{\omega}_{\p}}\int_{-\tilde{\omega}_{\p}-\epsilon}^{-\tilde{\omega}_{\p}+\epsilon}
		\frac{dk^0}{k^0+\tilde{\omega}_{\p}-i\gamma_B/(2\tilde{\omega}_{\p})}
		\nonumber \\ &+&
		\frac{\rho(\tilde{\omega}_{\p})}{2\tilde{\omega}_{\p}}\int_{\tilde{\omega}_{\p}-\epsilon}^{\tilde{\omega}_{\p}+\epsilon}
		\frac{dk^0}{k^0-\tilde{\omega}_{\p}+i\gamma_B/(2\tilde{\omega}_{\p})}
		 \nonumber \\ &=& 
		  \textrm{not enhanced}
		-\frac{2i\epsilon}{\gamma_B}\l[\rho(\tilde{\omega}_{\p})+\rho(-\tilde{\omega}_{\p})
		 \r]~,
		\eea
where $\epsilon$ identifies areas that are close enough to the poles in a way that $\epsilon\ll \gamma_B$, and ``not enhanced'' denotes the contribution that is not enhanced by the presence of the poles.
In this expression, the real part is explicitly canceled after integrating around the side peaks. 
The imaginary part of \eqref{Ft1} is non-vanishing. Since the profile function $\rho(k^0)$ is an even function of $k^0$, the final expression for the imaginary part becomes $2\epsilon\rho(\tilde{\omega}_{\p})/\gamma_B$. The resulting expression in \eqref{integralk0} is small and so we neglect it.

\begin{figure*}
\hspace*{-1cm}
\centering
\includegraphics[width=.95\textwidth]{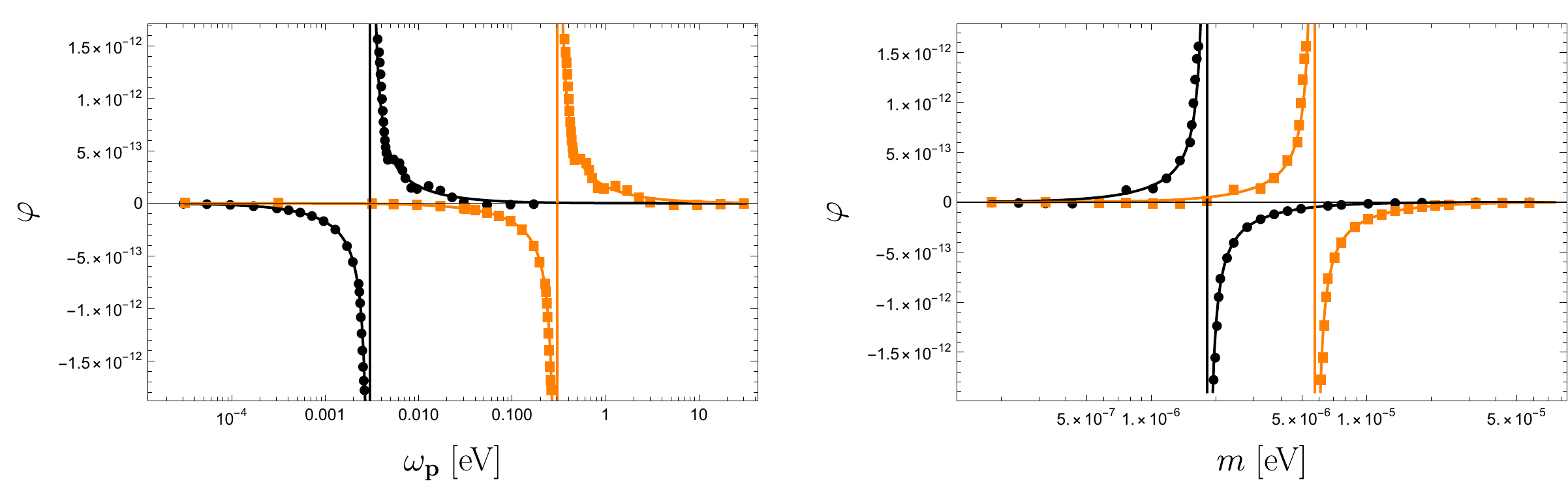}
\caption{\small Comparison of numerical calculation of $\varphi$ (markers), time integration on $\mathcal{F}$, with its analytical approximation (solid lines) given in Eq.~\eqref{int-profile-poles}, in terms of $\omega_\p$ (left panel) and $m$ (right panel). The black and the orange curves in the left panel correspond to $m_a=10^{-6}$\,eV and $m_a=10^{-5}$\,eV, respectively. In the right one, the black and orange curves represent $\omega_\p=10^{-2}$~eV and $\omega_\p=10^{-1}$~eV, respectively. The coupling constant and the magnetic field frequency supposed to be $g_{a\gamma\gamma}=10^{-10}$~GeV$^{-1}$ and $\omega_B=250$~kHz, respectively}. 
\label{numeric-vs-analytic}
\end{figure*}

We then proceed by providing an analytical expression for the integration around the peaks produced by the profile function. 
This approach provides a simple analytic form for the impact of the near resonance phenomenon on the phase and the amplitude of the output signal. 
To approximate the integration \eqref{intFt} around these peaks we note the presence of the $\rm sinc$ functions, emerging from inserting \eqref{P1} and \eqref{P2} into \eqref{Ft1}, and then in \eqref{intFt}. These $\rm sinc$ functions can be approximated with Dirac delta functions,
\beq
\frac{\sin(k^0-\omega)\tau_B/2}{(k^0-\omega)}\rightarrow\pi\delta(k^0-\omega)~,
\eeq
where $\omega$ is any of $\pm(\omega_\p\pm\omega_B)$, and where the approximation holds in the limit of $\omega \tau_B\gg1$
. Given that the integral becomes important around the location of peaks, one can evaluate the integration \eqref{intFt} as
\bea \label{alalytic-approx}
	\frac{\varphi}{G_a\omega_\p}&\simeq&
		\frac{\pi\tau_B/2}{\epsilon_++i\gamma_B}
	+\frac{\pi\tau_B/2}{\epsilon_-+i\gamma_B}	\label{int-profile-poles}~,
\eea
where
\beq
\epsilon_{\pm}=\omega_B(\omega_B\pm2\omega_\p)-m_a^2~.
\eeq
The imaginary part of \eqref{int-profile-poles} is small and it is neglected. 
In Fig~\eqref{numeric-vs-analytic}, we have plotted the real part of the expression \eqref{int-profile-poles} in terms of $m_a$ and $\omega_\p$ for $g_{ a\gamma\gamma}=10^{-10}\:\textrm{GeV}^{-1}$ and compared 
it with numerical results. The results show a perfect agreement between the analytical expression given by \eqref{int-profile-poles} and a full numerical computation. In the following, we evaluate Eqs.~\eqref{Langevins} numerically, and we obtain the region in the axion mass-coupling plane that can be probed by our proposed experiment.

\section{Sensitivity to the axion-photon coupling}

In this section, we estimate the sensitivity of our proposed experiment to the axion-photon coupling constant $g_{a\gamma\gamma}$ for a given ALP mass $m_a$.
The configuration of our setup leads to a phase and amplitude modification in the cavity polarization mode which is aligned with the external magnetic field and leaves the perpendicular polarization intact. We exploit this property for the detection of the axion-induced phase with a noise cancellation mechanism. To this end, both polarization modes are equally pumped with the same laser pulse and a Homodyne measurement is performed on the outgoing cavity field. The phase difference between the two polarizations thus signals the photon-axion interaction in the presence of the magnetic field. The perpendicular polarization mode (here $a_2$) is employed as a reliable phase reference. Note that since both cavity modes are pumped by the same optical drive and go through the same optical path inside the cavity, they share the laser phase noise as well as the thermal noise imposed from vibrations of the cavity walls. Therefore, the homodyne phase difference setup shown in Fig.~\eqref{polvec} cancels such joint phase fluctuation effects~\cite{Gardiner}.
In the homodyne detection, one is only interested in the phase difference
\begin{equation}
\phi_{12}(t) = \big|\mathrm{arg}\mean{a_1^{\rm out}(t)} -\mathrm{arg}\mean{a_2^{\rm out}(t)}\big|,
\end{equation}
where $\langle\cdot\rangle$ denotes the expectation value and $\mathrm{arg}(\alpha) \equiv \arctan(\Im\{\alpha\}/\Re\{\alpha\})$ gives the phase of $\alpha$, a complex number.
Here, the output cavity field is given by~\cite{Walls}
\begin{equation}
	a_i^{\rm out} = \sqrt{2\kappa}~a_i -a_i^{\rm in}.
\label{inout}
\end{equation}
In the following analysis, we employ a laser pulse with Gaussian envelope. Hence, the first moment of the input field operator for both polarizations is given by
\beq
\langle a_i^{\textrm{in}}(t)\rangle=\sqrt{\frac{P_L} {2\omega_{\textbf{p}}}}\,e^{-(t/\tau_{\!_L})^2}e^{-i\omega_{\bf p}t}~,
\label{infield}
\eeq
where $P_L$ denotes the input laser power and $\tau_{\!_L}$ is the length of the pulse. If long enough laser pulses ($\tau_{\!_L} \gg \tau,\tau_B$) are employed for pumping the cavity modes, the phase shift $\phi_{12}(t)$ keeps accumulating, while the oscillating background magnetic field is on. Then, the phase difference is maximum at the time $t=\tau_B/2$, when the external magnetic field is shut off.

The phase sensitivity in a single-shot optical scheme depends on the number of photons inside the probe and the uncertainty in its phase $\Delta\phi$. The larger number of photons, the higher becomes the minimum detectable phase. Nonetheless, there is a fundamental limitation in the width of an optical state in the corresponding phase space which imposes the standard quantum limit $\Delta\phi =1$.
The limit can be surpassed by modifying the optical field in a way that leads to the squeezing of its phase space representation along the phase direction, see Fig.~\eqref{scheme}.
Such states are created via nonlinear optical crystals and are usually parameterized by $r$, the squeezing parameter such that $\Delta\phi = e^{-r}$. Hence, the single-shot sensitivity is $\delta\phi=\frac{e^{-r}}{\sqrt{N_{\textrm{ph}}}}$, where $r=0$ gives the shot-noise limit \cite{gerry}.
One repeats the measurement $N_{\textrm{exp}}$ times to attain the higher sensitivity of $\delta\phi=\frac{e^{-r}}{\sqrt{N_{\textrm{ph}}N_{\textrm{exp}}}}$.
The number of photons inside the cavity depends on the laser power $P_L$, the cavity decay rate $\kappa$, and the pulse duration $\tau_{\!_L}$.
The number of photons contributing in the homodyne detection is given by $N_{\rm ph} = |\mean{a_1^{\rm out}}|^2\approx|\mean{a_2^{\rm out}}|^2$, where we have assumed that the effect of $\mathcal{F}_i$ on the amplitude of $a_1$ is negligible. This assumption is supported by our numerical analysis. We thus have
$$\delta\phi(t) = \frac{e^{-r}}{|\!\mean{a_1^{\rm out}(t)}\!|\sqrt{N_{\rm exp}}}.$$
Detection of any phase difference then requires $\phi_{12}(t) > \delta\phi (t)$. Therefore, by introducing the cavity phase-amplitude parameter $\Phi \equiv |\!\mean{a_1^{\rm out}(\frac{\tau_B}{2})}\!|\phi_{12}(\frac{\tau_B}{2})$ the detection condition becomes:
\begin{equation} \label{DetCon}
	\Phi > \frac{e^{-r}}{\sqrt{N_{\rm exp}}}.
\end{equation}

	
We evaluate $\Phi$ by numerically computing $\mean{a_1}$ and $\mean{a_2}$ from Eqs.~\eqref{Langevins} and \eqref{infield} and then plugging the solutions in Eq.~\eqref{inout}. For each value of ALP mass, we consider various laser frequencies to find the optimal one by which $\Phi$ gets maximized. It should be noticed that just $\omega_{\mathbf{p}}=n \textrm{FSR}, (n=1,2,3,...)$ are allowed for resonant laser frequencies.
The best signal is obtained when the propagator poles and the profile poles overlap indicating that the following resonance condition is met, 
\bea \label{Optimum}
m_a^2+\omega_{\mathbf{p}}^2\approx(\omega_{B}+\omega_{\mathbf{p}})^2~.
\eea
For each ALP mass when the laser frequency is set using Eq.\eqref{Optimum} , the numerical inspection likewise demonstrates an improvement. From now on, we choose the laser frequency $\omega_{\bf p}$ such that it satisfies the above resonance equation and yet is on resonance with one of the cavity normal frequencies.

\begin{figure}[t]
 \hspace*{-1cm}
 	\centering     
 	\includegraphics[width=.85\textwidth]{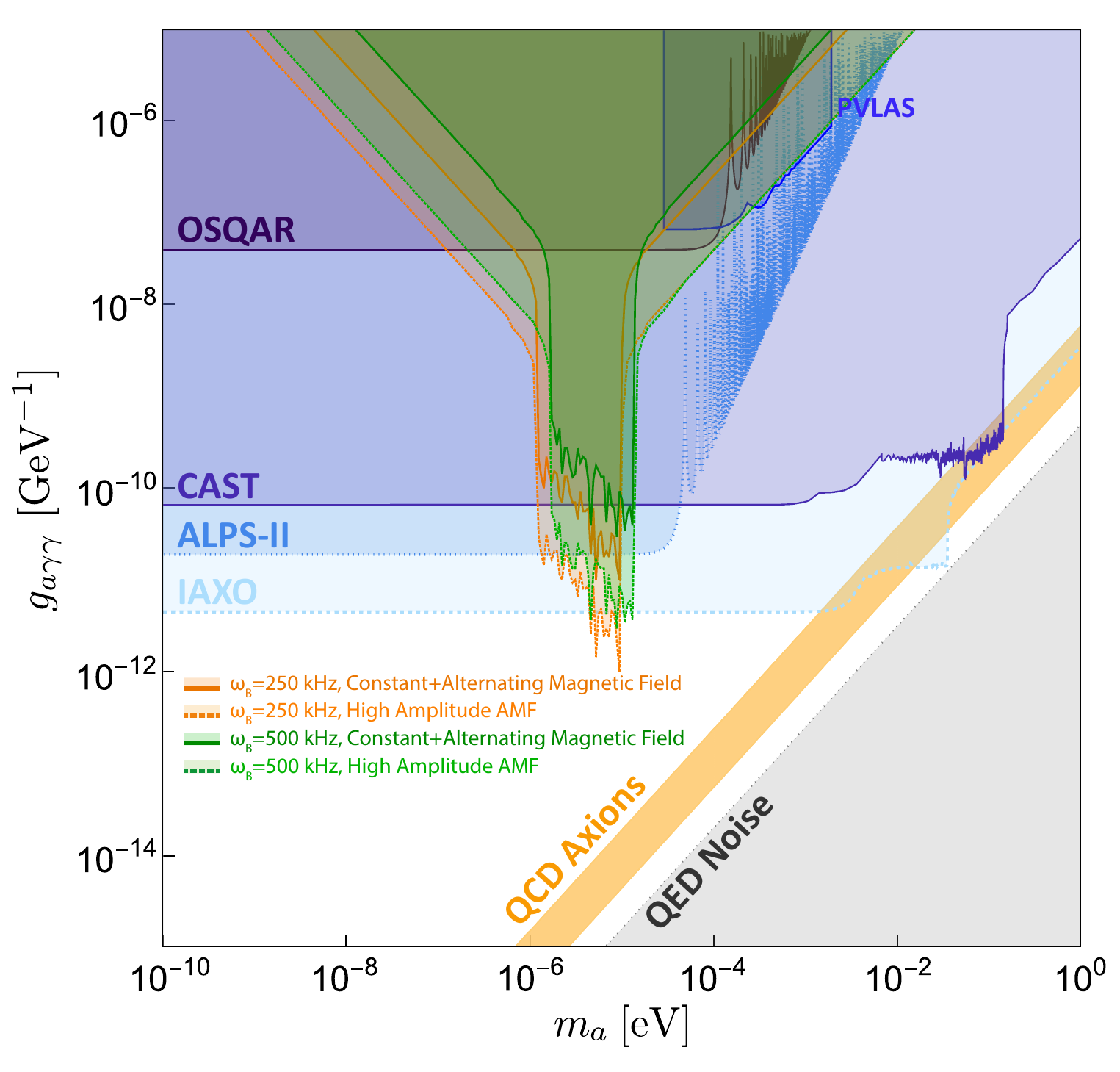}
 	\advance\leftskip-.4cm
 	\caption{\small 
 	Exclusion areas in the plane of ALP mass and axion-photon coupling constant. In the first scenario (dashed lines) we employ a high amplitude Alternating Magnetic Field (AMF) with $B_0=1\,\rm T$. In the second one (solid lines), we combine a constant magnetic field of $B_1=10\,\rm T$ with a moderate AMF of $B_0=10\,\rm mT$. The orange and green plots correspond to $\omega_B=250$\,kHz and $\omega_B=500$\,kHz,  respectively. The laser power assumed to be $100\:\rm mW$ and a squeezed light with the squeezing parameter of $r\approx1.75 ~\textrm{dB}$ is used. The gray shade is the region where QED-induced polarization-flip dominates that of axion-induced processes~\cite{Zarei:2019sva}.}
 	\label{exclusion}
 \end{figure}

 In the following analysis, a laser device is employed with a power of $P_L=100$ mW that pumps photons into a Fabry-Perot cavity with a length of $L_x=10$ cm and a finesse of $F=10^6$ thus the photon lifetime in the cavity is $\tau=\frac{L_xF}{\pi c}\approx\:106\:\mu\textrm{s}$. We consider a squeezed light with the squeezing parameter $r\approx1.75 ~\textrm{dB}$ which is fully compatible with the current technology \cite{andersen2016}. It is also worth remarking that in our analysis we assume that the interaction time scale is smaller than the photon lifetime $\tau$. Hence, 
 we choose the parameters such that the inequality $\tau_B<\tau$ always holds.  In order to obtain the optimal value of $\omega_\p$, Eq. \eqref{Optimum} is solved for each $m_a$ and the nearest integer multiple of FSR is picked up within the range of $0.03\:\textrm{eV}\leq\omega_\mathbf{p}\leq3\:\textrm{eV}$. Note that this restriction comes from the accuracy of the homodyne detection scheme. We use the analytical integration in Eq. \eqref{alalytic-approx} and numerically solve the expectation values from Langevin equations in Eqs.~\eqref{Langevins}. The answers are plugged in Eq. \eqref{inout} and then by applying Eq. \eqref{DetCon} the resulting exclusion region is found which is illustrated in Fig.~\eqref{exclusion}. We apply our scheme to the two aforementioned scenarios indicated in Section II subsections. In each scenario, there are two cases of magnetic field frequency. In one case the magnetic field frequency is $\omega_B=250\:\textrm{kHz}$ (which is demonstrated by orange plots in Fig.~\eqref{exclusion}) so with $n=3$, we have $\tau_B=(2n+1)\pi/\omega_B\approx\:94\:\mu\textrm{s}$ thus $\tau_B/\tau\simeq0.8$. In the other case, the magnetic field frequency is $\omega_B=500\:\textrm{kHz}$ (which is demonstrated by orange plots in Fig.~\eqref{exclusion}) so using $n=7$, we have $\tau_B/\tau\simeq0.9$. Therefore, in both cases, the time duration of the magnetic field is less than the photon lifetime inside the cavity. 
 	
In the first scenario based on subsection \ref{section:AMF}, a modest alternating magnetic field with $B_0 = 1\:\textrm{T}$ ~\cite{HofmannPat1985,CostaPat1987,TaylorPat,schwarzPat,christiansen2017} is taken into account. The exclusion region of this scenario is illustrated in Fig.~\eqref{exclusion} as dashed lines for $\omega_B=250\:\textrm{kHz}$ (orange one) and $\omega_B=500\:\textrm{kHz}$ (green one). These two cases demonstrate the exclusion region's sensitivity to magnetic field frequency. The lighter ALP values are covered by reducing the magnetic field frequency.

In the second scenario based on subsection \ref{section:CAMF}, a constant magnetic field of $B_1=$10 T is used \cite{CAST:2017uph}. Also, the cavity is exposed to an alternating magnetic field of $B_0=$10 mT. Once again, there are two cases in which the frequency of the alternating magnetic field is $\omega_B=250$\,kHz and $\omega_B=500$\,kHz that are shown in Fig.~\eqref{exclusion} with orange and green dashed lines, respectively. The main reason for the enhancement effect of this scenario is the profile functions in the third line at the right hand side of Eq.~\eqref{DCAC}. According to Eqs.~\eqref{P1} and \eqref{P2}, there are four sharp peaks due to the profile functions on $k_0=\pm\omega_{\mathbf{ p}}\pm\omega_{B}$. Unlike the first scenario, in the third line of Eq.~\eqref{DCAC} the sinc functions of are not quadratic so the enhancement effect reduces but still exists. Furthermore, the amplitude of a uniform magnetic field, $B_1$, makes the third line of Eq.~\eqref{DCAC} more effective than its fourth line in which the only effective amplitude of magnetic fields is the alternating one, $B_0$.

According to Eq.~\eqref{Optimum}, by employing cavity mode frequencies $0.03\:\textrm{eV}\leq\omega_\mathbf{p}\leq3\:\textrm{eV}$ the ALP mass range of $3.1\:\mu$eV to $44.4\:\mu$eV is covered by the scenarios considered in this work. 

In generating the exclusion regions, we have employed optical pulses with duration $\tau_{\!_L}=3\tau$ which is the optimal value found by inspection.
This constitutes the main experiment duration factor. By considering the optical pulse duration and the homodyne detection we take $\tau_{\rm exp}\sim 10\tau$ as an estimation for each run of the experiment. Taking into account the number of experiment repetition $N_{\rm exp}=10^6$, one finds $N_{\rm exp}\tau_{\rm exp} \approx 17\,{\rm min}$ for the time that takes for excluding/detecting each ALP mass. Hence, in the time span of a month about 2,500 mass values can be examined. These 2,500 points per month can accurately cover the target mass region ($3.1\:\mu$eV - $44.4\:\mu$eV) in a typical experiment lifetime which is about three years. Although in the other mass regions one can use the scheme, the resonant condition cannot be obtained and the results of detection decline as shown in Fig.\eqref{exclusion}. Note that only ALP mass values that match the resonance condition in our setup can be efficiently tested. Hence, for covering a small range of $m_a$ a cavity with tunable length should be employed.

In addition to producing promising results when compared to previous trials, the technique has the flexibility to increase sensitivity.
 When a clue of a faint signal in a restricted mass region is discovered by another experiment, this approach might be employed as a follow-up search. Since the limiting element of time is less effective in this circumstance, our system may be utilized to scan this restricted region for a comprehensive survey. Knowing the specific mass region allows us to expand the number of tests $N_\text{exp}$ and improve the experiment's sensitivity using Eq.\eqref{DetCon}. Eventually, by searching in a very limited mass range between $3.1\;\mu\textrm{eV}\leqslant m_a \leqslant 44.4\;\mu\textrm{eV}$, one may reach the QCD axion bar in Fig.\:\eqref{exclusion} with a larger number of tests in the same measurement time.

\section{Summary and Conclusion}

In this paper, we have studied the interaction of cavity photons with a time-dependent external magnetic field mediated by off-shell axion-like particles. Starting from the S-matrix, we first derived an effective interaction Hamiltonian describing this process.
Remarkably, we have found that the time-dependency of the external magnetic field gives rise to a significant enhancement in the coupling of the photon-axion interaction emerging in the effective Hamiltonian. We thus have proposed to take the advantage of this enhancement property in a novel tabletop experiment for the detection of possible ALPs in an unexplored parameter area of axion mass and coupling rate. 
Our scheme relies on the quantum assisted precision optical phase measurement.

By employing proper polarization, power, and duration laser drive pulses as well as setup geometry, we have engineered a noise cancellation mechanism where the laser phase noise and mirror vibrational noises are canceled out in the homodyne measurement. All the facilities used in our proposed experiment are accessible in near future. Using our scheme, it may be possible to detect ALPs with much greater sensitivity. As shown in Fig.\:\eqref{exclusion}, over a particle mass range $3.1\;\mu\textrm{eV}\leqslant m_a \leqslant 44.4\;\mu\textrm{eV}$ we can measure the phase difference down to a photon-ALPs coupling constant of $g_{a\gamma\gamma}\sim10^{-12}\:\text{GeV}^{-1}$. Additionally, if higher frequency oscillatory magnetic fields become available in the future, a larger area of the parameter space can be explored by our scheme.

\begin{acknowledgments}
		
M.Z. acknowledges financial support by the University of Padova under the MSCA Seal of Excellence @UniPD programme. M.Z. also thanks Hossein Ahmadvand at the Isfahan University of Technology for the experimental discussions.
		
\end{acknowledgments}

	
\appendix
\section{ALPs effective propagator }
\label{appendix:propagator}

\begin{figure} [t]
	\centering
	\includegraphics[width=\textwidth]{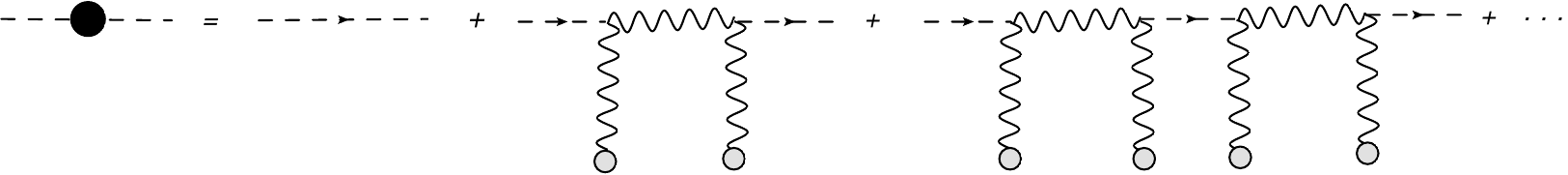}
	\caption{\small Axion self energy in the presence of a time-dependent magnetic field.}
	\label{axion-self}
\end{figure}

Now, we will focus on calculating the exact propagator $D_F(k^0)$. Fig.~\eqref{axion-self} shows the corrections to the propagator of ALPs. The effective propagator is given by 
\bea
	iD_F(k^0)&=&\frac{i}{\omega^2_\k-\p^2-m_0^2}+\frac{i}{\omega^2_\k-\p^2-m_0^2}\left(-i\Pi_B(k^0)\right)\frac{i}{\omega^2_\k-\p^2-m_0^2}+\cdot \cdot \cdot \nonumber\\ &=& \frac{i}{\omega^2_\k-\p^2-m_a^2-i\textrm{Im}\Pi_B(k^0)}~,
\eea
where $\Pi_B(k^0)$ denotes the sum of all one-particle irreducible (1PI) diagrams including diagrams with two external time-dependent magnetic field lines and $m_a$ is the physical axion mass. The imaginary part of $\Pi_B(k^0)$ can be identified as the conversion rate of axion into photon in the presence of the background magnetic field \cite{Zarei:2019sva}
	\beq
		\textrm{Im}\Pi_B(k^0)=k^0\Gamma_B(k^0)~.
	\eeq
Therefore, we can write the propagator as
\bea
	iD_F(k^0)=\frac{i}{\omega^2_\k-\p^2-m_a^2+ik^0\Gamma_B(k^0)}~,
\eea
\section{ALPs decay rate in the background of a time-dependent magnetic field} \label{appendix:decayrate}

In this section, we calculate the decay rate of axion in the presence of a time-dependent magnetic field. The amplitude for the axion-photon conversion process in the presence of the magnetic field is given by
\bea
\mathcal{T}_{fi}&=&\left<a,\mathbf{ k}\right|\frac{g_{ a\gamma\gamma}}{2}\int d^4x\, \epsilon^{\mu\nu\rho\sigma} \phi(x)\bar{F}_{\rho\sigma}(x)\partial_{\mu}A_{\nu}(x)\left|\gamma,\mathbf{p}\right>\\ \nonumber 
&=&-\frac{ig_{\gamma a} \sqrt{\omega_{\mathbf{p}}}}{ V\sqrt{2k^0}}
\int_{-\tau_B}^{\tau_B} dt e^{-i(\omega_{\mathbf{p}}-k^0) t}
\int_{V} dxd^2\x_\bot\,\sin(p_x x)e^{-ik_x x} e^{i(\mathbf{p}_\bot-\mathbf{k}_\bot)\cdot \mathbf{x}_\bot}\bm{\varepsilon}(\mathbf{ p})\cdot\mathbf{B} (t)~.
\eea	
Now, for a harmonic magnetic field $\mathbf{B}(t)= \mathbf{B}_{0}\cos{(\omega_Bt)}$, the conversion amplitude transforms to the form
\bea
\mathcal{T}_{fi}&=&-\frac{
	ig_{a\gamma\gamma} \sqrt{\omega_{\mathbf{p}}}\bm{\varepsilon}_{s}(\mathbf{ p})\cdot\mathbf{B}_{0}}{ V \sqrt{2k^0}}
\int_{-T/2}^{T/2} dt\, e^{-i(\omega_{\mathbf{p}}-k^0) t}\cos\omega_Bt
\int_{V} dx\,d^2\x_\bot\sin(p_x x) e^{-ik_x x} e^{i(\mathbf{p}_\bot-\mathbf{k}_\bot)\cdot \mathbf{x}_\bot}~.\nonumber\\
\eea
The conversion probability per unit time is given by
\bea
w=\frac{|\mathcal{T}_{fi}|^{2}}{T}=\frac{g^2_{ a\gamma\gamma}   \omega_{\mathbf{p}}}{8L_yL_z\tau_B k^0 } |\mathbf{B}_{0}|^2P_1^2(k^0)\,\,(2\pi)^2 \delta^2(\mathbf{p}_\bot-\mathbf{k}_\bot)\,\delta_{p_x,k_x}  ~,
\eea 
in which we have ignored the boundary conditions for simplicity and also in a good approximation we have set $T=\tau_B$.	The decay rate is given by
\bea\label{Gmmm}
\Gamma_B(k^0)=\sum_{p_x}\int w \frac{L_yL_zd^2\p_\bot}{(2\pi)^2}=\frac{g^2_{ a\gamma\gamma} \omega_\p }{8k^0} |\mathbf{B}_{0}|^2\frac{P_1^2(k^0)}{\tau_B}~.
\eea
Approximating $P_1^2(k^0)/\tau_B\approx \tau_B/4$, we find a maximum value for the decay rate as
\bea \label{barGmmm}
\Gamma^{\textrm{max}}_B=\frac{g^2_{ a\gamma\gamma} \omega_\p }{ 32\, k^0} |\mathbf{B}_{0}|^2 \tau_B
~.
\eea

\section{Interaction Hamiltonian Calculation}\label{appendix:interaction}
\begin{figure*}[t]
	\centering
	\captionsetup[subfigure]{oneside,margin={.32cm,0cm}}
	\begin{subfigure}{0.47\textwidth}
		\hspace*{-.8cm}
		\includegraphics[width=\textwidth]{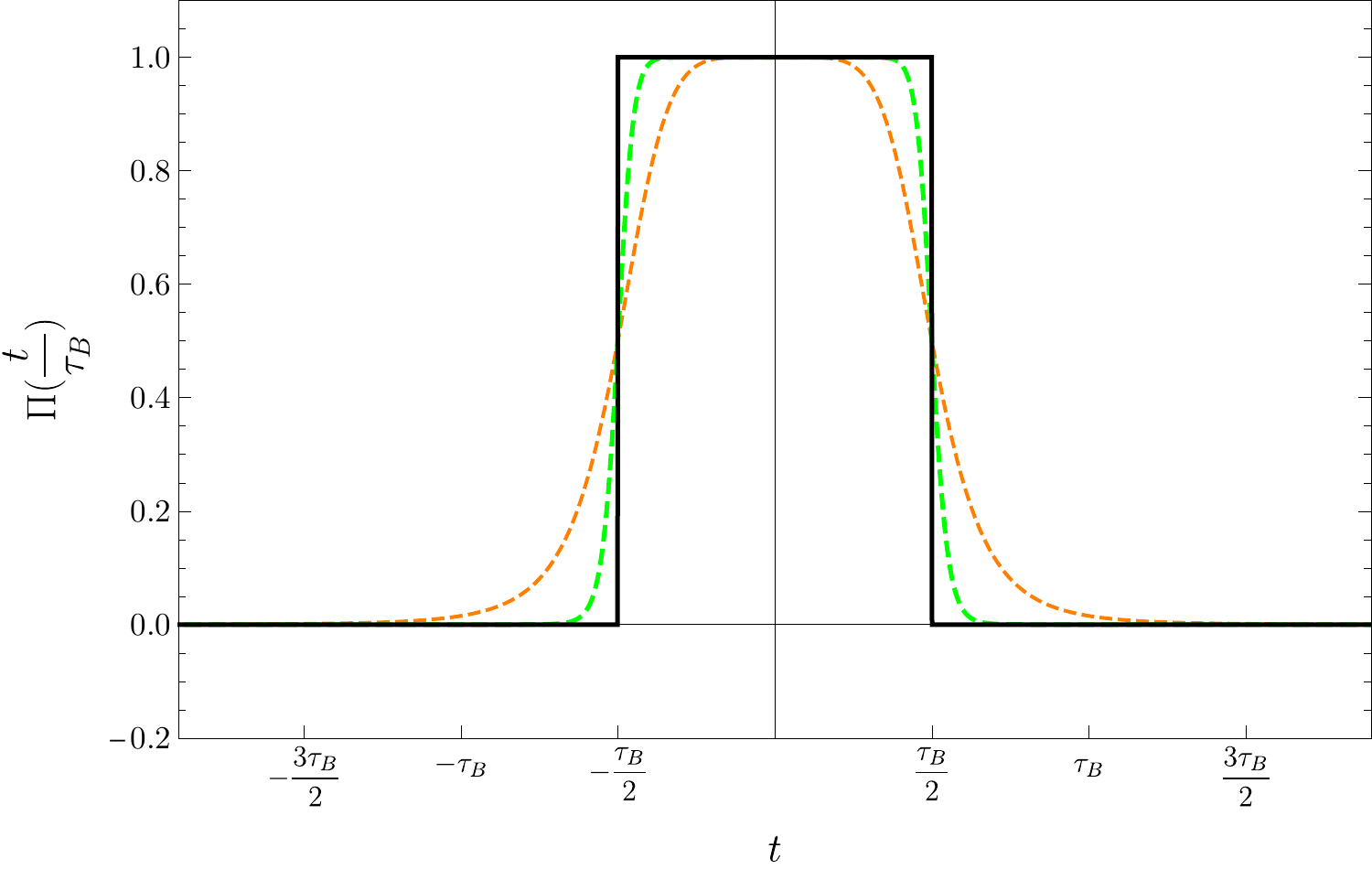}
		\caption{}
		\label{fig:Rect}
	\end{subfigure}
	\quad
	\captionsetup[subfigure]{oneside,margin={.7cm,0cm}}
	\begin{subfigure}{0.47\textwidth}
		\includegraphics[width=\textwidth]{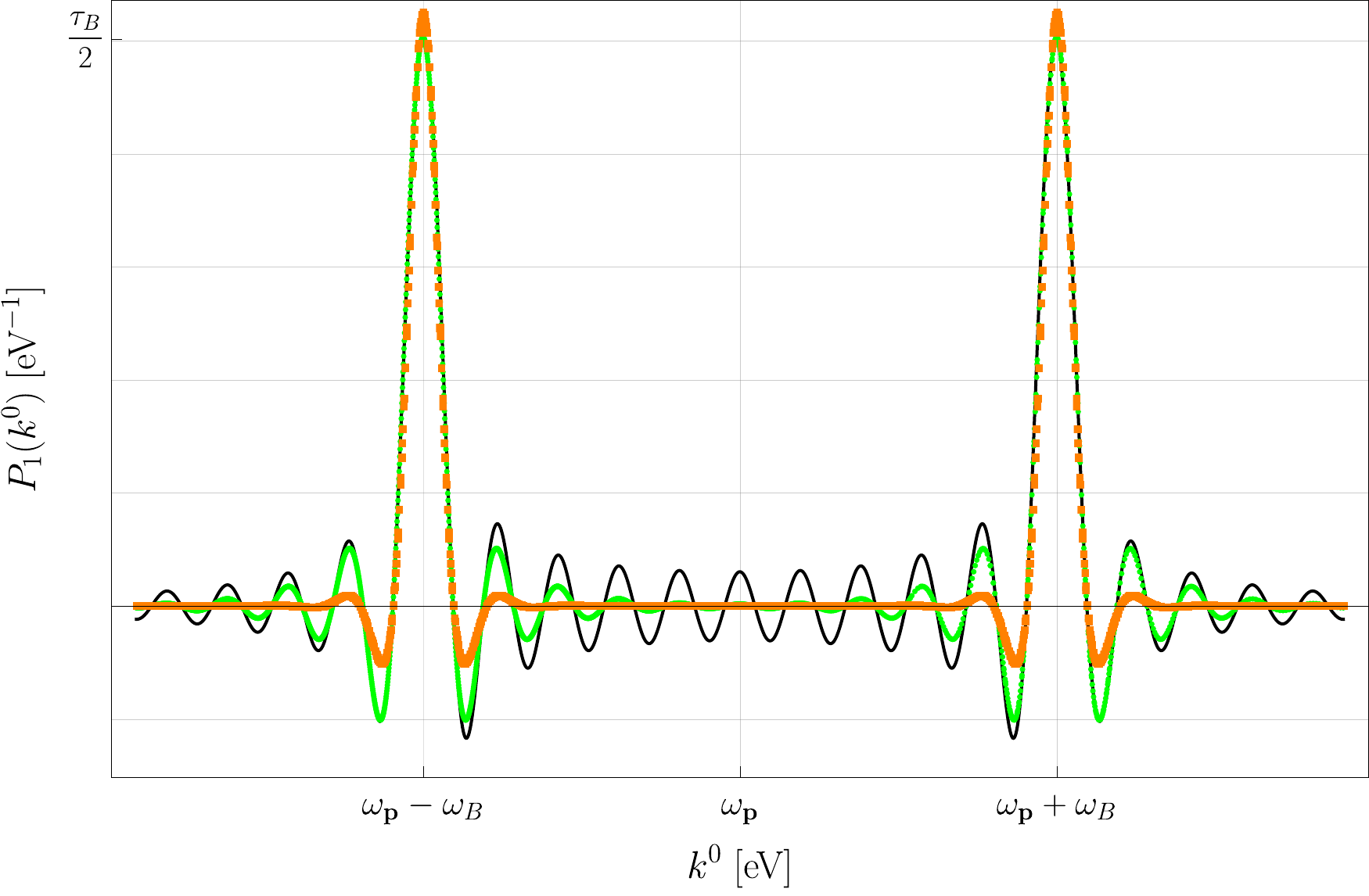}
		\caption{} \label{fig:Profile1}
	\end{subfigure}
	\captionsetup{singlelinecheck=false}
	\caption{ \small (a) Comparison of the rectangular function $\Pi(t)$ (solid black) with its approximation in relation \eqref{rec}. Dashed orange and long-dashed green curves belong to $n=6$ and $n=20$, respectively. (b) Effect of approximating rectangular function  on first profile in \eqref{P1}. The solid black line is $P_1(k^0)$ with the exact rectangular function. Orange and green dotted curves belong to numerical calculation of $P_1(k^0)$ in which rectangular function in \eqref{rec} is approximated  with $n=6$ and $n=20$, respectively. } \label{newprofile}
\end{figure*}

This appendix contains a detailed calculation of the interaction Hamiltonian $H(t)$. In order to see the effect of magnetic field oscillation on the interaction, an optimal condition is obtained when the $\cos{\omega_B t}$ term in Eq.~\eqref{bg} oscillates several times in the $\tau_B$ interval. 
This condition is met for $1/\omega_B\ll \tau_B$. Moreover, since we want to make sure that the magnetic field triggers from zero and increases gradually with time, $\tau_B$ should be as $\tau_B=(2n+1)\pi/\omega_B$ in which $n$ is a natural number that must be $n\gg1$ to satisfy the condition $1/\omega_B\ll \tau_B$. Inserting \eqref{bg} in the interaction Hamiltonian \eqref{H}, carrying out the integration over $\bm{x}_{\bot}$, $\bm{x}'_{\bot}$ and $\mathbf{k}_\bot$ yields
\bea \label{Hint1-1}
H(t)&=&\frac{g^2_{ a\gamma\gamma}}{2 V}\,\cos(\omega_Bt)\,\Pi\l(\frac{t}{\tau_B}\r)
\,\sum_{s,s'}\sum_{\p ,\p'}\sqrt{\omega_\mathbf{p}\omega_\mathbf{p'}}\Big[\bm{\varepsilon}_{s'}^*(\p')\cdot\mathbf{B}_0\Big]\Big[\bm{\varepsilon}_{s}(\p)\cdot\mathbf{B}_0\Big]a_{s'}^{\dagger}(\p')a_{s}(\p) \nonumber \\ &\times &
\int dt'dx'dx\,\frac{dk^0}{2\pi}\frac{dk_x}{2\pi}\,\cos(\omega_Bt')\,\Pi\l(\frac{t'}{\tau_B}\r)\frac{1}{{k^0}^2-m_a^2-k_x^2-\mathbf{p}_\bot^2+ik^0\Gamma_B(k^0)}(2\pi)^2
\delta^2(\mathbf{p}_\bot-\mathbf{p}'_\bot)
\nonumber\\
&\times&
\left[
\sin(p'_{x} x)\sin(p_{x} x')
e^{-ik_x x'}
e^{i k_x x }
e^{-i(-k^0+\omega_{\mathbf{p}})t'}
e^{-i(k^0-\omega_{\mathbf{p}'})t}
\right.\nonumber\\
&+&\left.
\sin(p'_{x} x')\sin(p_{x} x)
e^{-ik_x x' }
e^{i k_x x }
e^{-i(-k^0-\omega_{\p'})t'}
e^{-i(k^0+\omega_{\p})t}\right]~.
\eea
By replacing the Dirac delta function with the discrete Kronecker delta function as $(2\pi)^2\delta^2(\mathbf{p}_\bot-\mathbf{p}'_\bot)=L_yL_z\delta^2_{\mathbf{p}_\bot,\mathbf{p}'_\bot} $ and  after integrating over $t'$ we find
\bea \label{Hint1-2}
H(t)&=&\frac{g^2_{ a\gamma\gamma}}{2 L_x}
\,\cos(\omega_Bt)\,\Pi\l(\frac{t}{\tau_B}\r)\sum_{s,s'}\sum_{\p,p'_x}\sqrt{\omega_\mathbf{p}\omega_\mathbf{p'}}\Big[\bm{\varepsilon}_{s'}^*(\p')\cdot\mathbf{B}_0\Big]\Big[\bm{\varepsilon}_{s}(\p)\cdot\mathbf{B}_0\Big]a_{s'}^{\dagger}(\p')a_{s}(\p) \nonumber \\ &\times &
\int \,\frac{dk^0}{2\pi}\frac{dk_x}{2\pi}\,\frac{1}{{k^0}^2-m_a^2-k_x^2-\mathbf{p}_\bot^2+ik^0\Gamma_B(k^0)}
\nonumber \\ &
\times &
\left[
\int_{0}^{L_x} dx\sin(p'_{x} x)e^{-ik_x  x }
\int_{0}^{L_x} dx'\sin(p_{x} x')e^{ik_{x}x' }
P_1(k^0) 
e^{-i(k^0-\omega_{\mathbf{p}'})t}\right.
\nonumber\\
&&\left.+
\int_{0}^{L_x} dx \sin(p_{x} x)e^{-ik_x  x }
\int_{0}^{L_x} dx'\sin(p'_{x} x')e^{ik_{x}x' }
P_2(k^0)
e^{-i(k^0+\omega_{\p})t}
\right]~,
\eea
where
\bea \label{appendixP1}
P_1(k^0)=\int_{-\infty}^{\infty}  dt' \cos(\omega_Bt')\,\Pi\l(\frac{t'}{\tau_B}\r)e^{-i(-k^0+\omega_{\p})t'}=\frac{\sin{\left(\Delta_1\tau_B/2\right)}}{\Delta_1}+\frac{\sin{\left( \Delta_2 \tau_B/2\right)}}{\Delta_2}~,
\eea
and
\bea  \label{appendixP2}
P_2(k^0)=\int_{-\infty}^{\infty}  dt' \cos(\omega_B t')\,\Pi\l(\frac{t'}{\tau_B}\r)e^{-i(-k^0-\omega_{\p})t'}=
\frac{\sin{\left(\Delta_3\tau_B/2\right)}}{\Delta_3}+\frac{\sin{\left( \Delta_4 \tau_B/2\right)}}{\Delta_4}~,
\eea
with $\Delta_1=k^0+\omega_B-\omega_{\mathbf{p}}$, $\Delta_2=k^0-\omega_B-\omega_{\mathbf{p}}$
, $\Delta_3=k^0+\omega_B+\omega_{\mathbf{p}}$, and $\Delta_4=k^0-\omega_B+\omega_{\mathbf{p}}$. 
The $sinc$ functions in \eqref{appendixP1} and \eqref{appendixP2} appear due to the rectangular form that we have considered for the magnetic field pulse, $\Pi(t/\tau_B)$. In appendix \ref{appendix:approximation}, we replace $\Pi(t/\tau_B)$ with a smoother function and compute \eqref{appendixP1} numerically. Our numerical investigations assure that considering a rectangular profile with smoothened edges (see dashed curves in Fig.~\eqref{fig:Rect}) does not affect the results obtained below, using Eqs.~\eqref{appendixP1} and \eqref{appendixP2}.

Assuming that the wavelength of the photons inside the cavity is much smaller than the length of the cavity, 
$\lambda/L_x \ll 1 $, one can approximately write the results of the integration over $x$ and $x'$ in terms of the Dirac delta function 
\bea \label{xintegral}
\int_{0}^{L_x} dx\sin(p_x x)e^{-ik_x  x }\int_{0}^{L_x} dx'\sin(p'_x x')e^{ik_x  x' }&\to&\frac{\pi^2}{4}\big[\delta(k_x+p_x)-\delta(k_x-p_x)\big
]\big[\delta(k_x+p'_x)
\nonumber\\
&& \:\:\:\:\:\:\:\:\:\:\:\:-\:
\delta(k_x-p'_x)\big]~.
\eea
Substituting \eqref{xintegral} into \eqref{Hint1-2} and integrating over $k_x$ we obtain 
\bea
H(t)&=&\frac{g^2_{ a\gamma\gamma}}{2 L_x}
\,\cos(\omega_Bt)\,\Pi\l(\frac{t}{\tau_B}\r)\sum_{s,s'}\sum_{\p\,p'_x} \sqrt{\omega_\mathbf{p}\omega_\mathbf{p'}}\Big[\bm{\varepsilon}_{s'}^*(\p')\cdot\mathbf{B}_0\Big]\Big[\bm{\varepsilon}_{s}(\p)\cdot\mathbf{B}_0\Big]a^{\dag}_{s'}(\textbf{p}')a_{s}(\p) \nonumber \\ &\times &
\int_{-\infty}^\infty \frac{dk^0}{2\pi}\,\frac{1}{{k^0}^2-m_a^2-p_x^2-\mathbf{p}_\bot^2+ik^0\Gamma_B(k^0)}\frac{\pi}{8}
\left(\delta(p_x-p'_x)-\delta(p_x+p'_x)-\delta(p'_x+p_x)\right.\nonumber \\ &~&~~~~~+\left. \delta(p'_x-p_x)\right)
\times 
\left[
P_1(k^0)
e^{-i(k^0-\omega_{\mathbf{p}'})t}
+
P_2(k^0)
e^{-i(k^0+\omega_{\p})t}
\right]~.
\eea
Exploiting the discrete form of delta function relation $2\pi\delta(p_x-p'_x)=L_x\delta_{p_x,p'_x} $, we get
\bea
H(t)&=& \frac{g^2_{ a\gamma\gamma}}{2 L_x}
\,\cos(\omega_Bt)\,\Pi\l(\frac{t}{\tau_B}\r)\sum_{s,s'}\sum_{\p,\,p'_x} \sqrt{\omega_\mathbf{p}\omega_\mathbf{p'}}\Big[\bm{\varepsilon}_{s'}^*(\p')\cdot\mathbf{B}_0\Big]\Big[\bm{\varepsilon}_{s}(\p)\cdot\mathbf{B}_0\Big]a^{\dag}_{s'}(\textbf{p}')a_{s}(\p) \nonumber \\ &\times &
\int_{-\infty}^\infty \frac{dk^0}{2\pi}\,\frac{1}{{k^0}^2-m_a^2-p_x^2-\mathbf{p}_\bot^2+ik^0\Gamma_B(k^0)}\frac{L_x}{16}
\left(
\delta_{p_x,p'_x}-\delta_{p_x,-p'_x}
-
\delta_{-p_x,p'_x}+\delta_{p_x,p'_x}
\right)
\nonumber \\ &
\times &
\left[P_1(k^0)e^{-it(k^0-\omega_{\mathbf{ p}'})}
+P_2(k^0)e^{-it(k^0+\omega_{\mathbf{ p}})}
\right]~,
\eea
which leads to
\bea \label{appendix:Hint1-3}
H(t)=\sum_{s,s'}\sum_{\p}\mathcal{F}(t)\,\l[\bm{\varepsilon}_{s'}^*(\p)\cdot\hat{\mathbf{b}}\r]
\l[\bm{\varepsilon}_{s}(\p)\cdot\hat{\mathbf{b}}\r]a_{s'}^{\dag}(\p) a_s(\p)~,
\eea 
in which $\hat{\mathbf{b}}$ is the direction of magnetic field and $\mathcal{F}(t)$ is
\bea \label{appendix:Ft1}
\mathcal{F}(t)&=&G_a\,\cos(\omega_B t)\,\Pi\l(\frac{t}{\tau_B}\r)\,\omega_{\mathbf{p}}\,\int_{-\infty}^\infty dk^0
\frac{1}{{k^0}^2-\tilde{\omega}_{\mathbf{p}}^2+ik^0\Gamma_B(k^0)}\l[e^{-it(k^0-\omega_{\mathbf{ p}})}
P_1(k^0)
\right. \nonumber \\ &+& \left. e^{-it(k^0+\omega_{\mathbf{ p}})}
P_2(k^0)\r]~,
\eea
where $G_a= \frac{g^2_{ a\gamma\gamma}B_0^2}{32\pi} $, and $\tilde{\omega}^2_{\mathbf{p}}=\omega^2_{\mathbf{p}}+m_a^2$.


\section{A more feasible time-dependent magnetic field }\label{appendix:approximation}

In section II, we analyzed the effects of a time-dependent magnetic field with a rectangular profile function $\Pi(t/\tau_B)$. We investigate a more practical profile that may be more feasible experimentally. To this end, we examine the following approximation of $\Pi(t)$ 
\beq \label{rec}
\Pi(t)=
\lim_{n \to \infty}\frac{1}{(2t)^n+1} ~,
\eeq
where $n$ is a positive integer. It follows that the rectangular function $\Pi(t)$
can be approximated with high accuracy by taking sufficiently large integer $n$. The Fig.~\eqref{fig:Rect} shows $\Pi(t/\tau_B)$ in comparison with the approximation function \eqref{rec} for $n=6$ and $n=20$. The larger the value of $n$, the more accurate agreement with the $\Pi(t/\tau_B)$ function. 
As discussed in the text, due to the presence of the $\Pi(t/\tau_B)$ function and after integrating over time in equations \eqref{P1} and \eqref{P2}, the $ P_1(k^0)$ and $P_2(k^0)$ are obtained in terms of the $sinc$ functions. Here, we replace the $\Pi(t/\tau_B)$ with the approximation expression \eqref{rec} and once again calculated \eqref{P1} and \eqref{P2} numerically.
In Fig.~\eqref{newprofile} we have displayed the results for $P_1(k^0)$ (a similar result is obtained for $P_2(k^0)$). 
As this figure shows, if we replace the rectangular function with the approximate function \eqref{rec}, the main shape of the peaks will not change. Therefore, we do not expect the main results obtained in the text to be affected by the values of $n$.

\pagebreak

\end{document}